\documentclass[10pt,a4paper,onecolumn]{article}

\usepackage[journal=jpclcd]{achemso}
\setkeys{acs}{articletitle = true}
\usepackage[T1]{fontenc}       
\usepackage{graphicx}
\usepackage{dblfloatfix}
\usepackage{placeins}
\usepackage{authblk}
\usepackage{a4wide}
\usepackage{setspace}
\usepackage[margin=1in]{geometry}
\usepackage{physics}
\usepackage{color}
\usepackage{bm}
\usepackage{booktabs, tabularx} 
\usepackage{amssymb}
\usepackage{url}
\usepackage[hidelinks]{hyperref}
\newcolumntype{C}{>{\centering\arraybackslash}X}

\author[1]{Amitav Sahu \thanks{amitav.sahu@chem.lu.se}}
\author[1]{T$\ddot{\text{o}}$nu Pullerits \thanks{tonu.pullerits@chem.lu.se}}
\affil[1]{Division of Chemical Physics and NanoLund, Lund University, Box 124, 221 00 Lund, Sweden}

\newcommand*{\si}[1]{\textcolor{black}{ #1}}
\newcommand*{\as}[1]{\textcolor{black}{ #1}}
\begin{document}

\title {Disorder-Induced Eigenstate Localization of Molecular Polaritons Despite Spectroscopic Strong Coupling}

\maketitle

\begin{abstract}
Molecular polaritons are hybrid light--matter states whose collective character is associated with a molecular excitation extended across many molecules. However, clearly resolved polariton peaks do not by themselves establish that the corresponding eigenstates retain this molecular delocalization. Here, we use a disordered Tavis--Cummings model to distinguish spectral signatures of strong coupling from eigenstate-level molecular delocalization of electronic polaritons. Static energetic disorder can strongly reduce the molecular extent of optically visible polariton states while resolved polariton peaks persist and their apparent separation can even increase. We quantify this crossover using complementary molecular localization measures applied to individual optically visible eigenstates and the photon-weighted bright sector. Finite-size scaling shows a stable large-ensemble crossover; for Gaussian disorder, retaining approximately half of the homogeneous molecular extent requires a collective coupling roughly seven times the Gaussian disorder standard deviation $\sigma$, although the numerical scale depends on the disorder distribution. Including cavity and molecular linewidths through a non-Hermitian Hamiltonian reveals a contrasting spectral and eigenstate response: finite linewidths reduce spectral resolvability, while cavity--molecule linewidth imbalance shifts the localization crossover to larger static disorder by suppressing the disproportionate contribution of near-resonant molecular transitions. Spectroscopic strong coupling and eigenstate-level molecular delocalization therefore characterize distinct properties and need not evolve in parallel under disorder and dissipation.
\end{abstract}

\section{\label{sec:Intro}Introduction}

Strong coupling between molecular excitations and confined optical fields gives rise to hybrid light--matter quasiparticles known as molecular polaritons. In this regime, the optical response is governed by collective eigenstates rather than by independent molecular and photonic excitations, and the corresponding normal mode splitting increases with the total molecular oscillator strength participating in the cavity mode\cite{Dicke1954,Tavis1968,Weisbuch1992,Ebbesen2016}. As a result, collective strong coupling can be achieved at experimentally accessible molecular concentrations even when the coupling of an individual molecule to the cavity field is weak. Molecular polaritons have attracted considerable interest because strong light--matter coupling has been associated with modified photophysics, energy transfer, charge transport, relaxation pathways, and chemical reactivity \cite{Hutchison2012,Orgiu2015,Thomas2016,Ebbesen2016,Ribeiro2018,Feist2018,GarciaVidal2021,LiNitzan2022,Arkajit2023Rev}. A key assumption underlying many of these interpretations is that the cavity couples to a collective bright molecular state, allowing the excitation to be distributed over many molecules rather than confined to a small subset of the ensemble.

The ideal collective picture is complicated in realistic molecular systems by energetic disorder and dissipation. Static disorder can arise from structural heterogeneity, molecular packing, orientational variations, and fluctuations of the surrounding environment, and is well known to influence exciton localization, optical broadening, and transport in molecular aggregates and organic materials\cite{Knapp1984,Bassler1993,Spano2010,Cho2005JPCB,Scholes2006,HestandSpano2018,Lev2022}. Molecular excitations also undergo homogeneous broadening through vibrational relaxation, pure dephasing, and non-radiative decay\cite{Mukamel1995,MayKuhn2011}, while optical cavities add photon loss through imperfect mirrors and radiative leakage\cite{GardinerCollett1985,Carmichael1993}. Molecular polaritons are therefore finite-lifetime states whose spectral properties and molecular extent are determined by a competition between collective coupling, energetic disorder, and dissipation.

The role of disorder in molecular polaritons has been examined from several complementary perspectives, and the resulting picture is not one of a single effect. Early studies of disordered microcavities showed that inhomogeneous broadening and disorder-induced scattering modify polariton linewidths and spectral visibility\cite{Weisbuch1992,Houdre1996,Whittaker1998}. Later theoretical work on molecular and organic cavities demonstrated that static disorder can redistribute oscillator strength, broaden polariton resonances, and convert nominally dark states into weakly bright, or ``gray'', states\cite{Michetti2005,Litinskaya2006,Grochol2008,Diniz2011,Cwik2016}. The disordered Tavis--Cummings model itself has been analyzed in considerable detail, including its eigenvalue structure and Fano-like resonances,\cite{Zeb2022PRA} as well as the disorder-induced restructuring of the polaritonic and dark-state manifolds\cite{Sebastian2022}. Beyond the single-mode description, disorder has been shown to influence polariton localization and dispersion in multimode microcavities\cite{Engelhardt2023PRL,Jianshu2022PRB} and to renormalize the polariton group velocity\cite{Aroeira2025Arxiv}. Importantly, these consequences are not uniformly detrimental. Disorder-activated gray states have been implicated in polariton-enhanced free-charge generation through a second-hybridization mechanism\cite{Wu2022JCP}, and disorder has been predicted to \emph{enhance} exciton transport and the quantum information spreading in cavity-coupled ensembles\cite{Wu2024Arxiv}. Further, static and dynamic disorder have been shown to act in contrasting ways on dark-state populations\cite{Catuto2025JCP}, and the interplay of disorder with phonon timescales controls dark-state activation in the thermodynamic limit\cite{Qiang2026Arxiv}. Disorder may additionally produce spectral splittings that need to be distinguished from genuine collective Rabi splitting\cite{Li2025JPCL}, while collective polaritonic effects on chemical dynamics can be suppressed by the same disorder\cite{Juan2024}. Recent experiments have also linked polariton delocalization to the nonlinear optical response of vibrational polaritons\cite{Xiang2019}, to ultrafast energy delocalization between spatially separated molecular layers\cite{Russo2024}, and to the coherent propagation of organic exciton--polaritons through delocalized dark states\cite{Pandya2021}. Taken together, these results show that disorder can affect spectral structure, eigenstate localization, transport, and dynamics in distinct ways. A statement about the survival of ``collective character'' must therefore specify which property or observable is being considered.

Despite these advances, a more specific question remains: when does an optically visible molecular polariton retain an extended molecular component in the presence of both static energetic disorder and finite linewidths? Recent studies have emphasized that spectroscopic strong coupling is not, by itself, a reliable proxy for molecular delocalization\cite{Sebastian2022,Wei2025ChemSci,Musser2024Expt}, and that disorder can substantially modify both linear optical observables and polariton-influenced chemical dynamics\cite{Joel2024NanoPhotonics,Juan2024}. Existing quantitative delocalization criteria\cite{Wei2025ChemSci}, however, have focused primarily on static energetic disorder, and have not explicitly examined how finite cavity and molecular linewidths modify the relation between spectral visibility and eigenstate localization. The relevant question is therefore not how large the Rabi splitting must be to produce resolvable polariton peaks, but how large the collective coupling must be to preserve an extended molecular component in the presence of energetic disorder and finite cavity and molecular linewidths. A second and complementary line of work should be distinguished from this one at the outset. Microscopic cavity quantum-electrodynamical treatments that go beyond effective few-level models resolve the modification of the molecular electronic structure itself under collective strong coupling, including local polarization of the individual molecules\cite{Sidler2021JPCL,Sidler2024JPCL,Horak2025PRR,Castagnola2025PRX,KochRubioJCP2021AbInitioStudy}. Those approaches ask what each molecule becomes inside the cavity; the present work asks how the molecular component of a single hybrid excitation is distributed among the members of a disordered ensemble, with the molecules themselves held fixed. The criterion developed here should therefore be understood specifically as a measure of \emph{eigenstate-level molecular delocalization}, rather than as a general criterion for polariton-modified transport, dynamics, or chemical reactivity.

Here we address this question using a disordered Tavis--Cummings model for an electronic molecular ensemble coupled to a single cavity mode. We define the model and its scope in Secs.~II.A and II.B. We examine the disorder-modified polariton spectrum in Sec.~II.C and quantify the corresponding loss of molecular delocalization in Sec.~II.D using three complementary measures of the molecular eigenvectors. The crossover is examined both at the individual-eigenstate and photon-weighted bright sector levels, including its dependence on ensemble size and on the form of the energetic disorder. In Sec.~II.E we extend the model to an effective non-Hermitian Hamiltonian and show that increasing cavity--molecule linewidth imbalance can shift the eigenstate-localization crossover toward larger static disorder even while spectral resolvability decreases. Section~II.F discusses the experimental implications, and Sec.~III summarizes the main conclusions.

\section{Theory, Results, and Discussion}

\subsection{Model Hamiltonian and Single-Excitation Basis}
\begin{figure}[ht!]
	\centering
	\includegraphics[width=3.5 in]{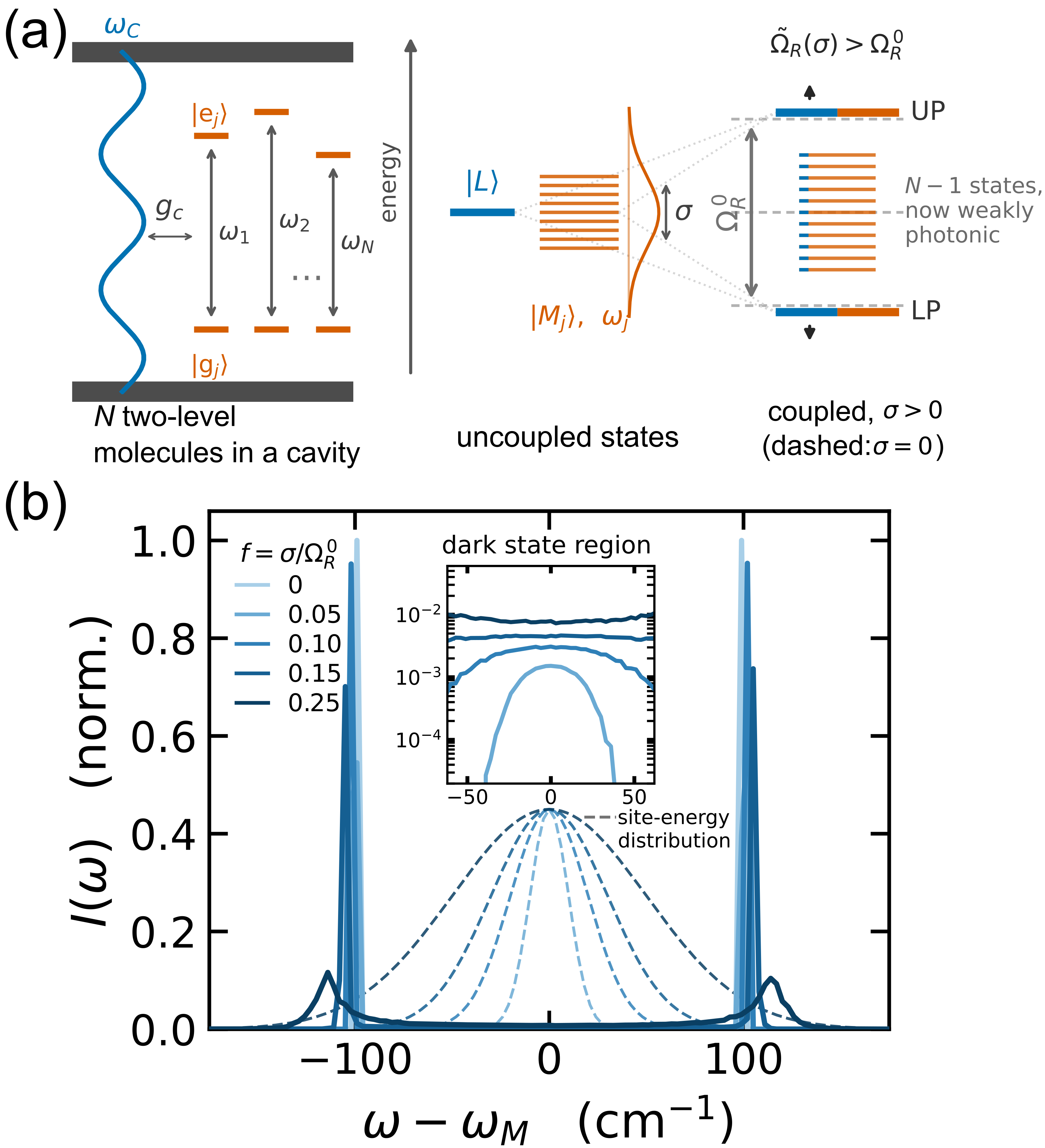}
	\caption{\footnotesize (a) Schematic representation of a cavity-coupled molecular ensemble. 
		$N$ molecules are modeled as two-level systems with ground state $|{\rm g}_j\rangle$, excited state $|{\rm e}_j\rangle$, and transition energy $\omega_j$ drawn from a Gaussian distribution of standard deviation $\sigma$ centered at $\omega_M$. The molecules are collectively coupled, with single-molecule coupling strength $g_c$, to a single-mode optical cavity of resonance energy $\omega_C$. The left column shows the uncoupled single-excitation basis, while the right column shows the coupled eigenstates at $\sigma>0$. Dashed lines mark the homogeneous ($\sigma=0$) polariton positions, separated at zero detuning by the bare collective coupling $\Omega_R^{0}=2g_c\sqrt{N}$, and the outward arrows indicate the increase of the observed spectral splitting $\widetilde{\Omega}_R$ with static disorder. The $N-1$ states that are degenerate and dark at $\sigma=0$ spread into a band of predominantly molecular character (orange) and acquire weak photonic character (blue). (b) Disorder-averaged cavity-probed spectral visibility for increasing static energetic disorder, expressed as the normalized disorder strength $f=\sigma/\Omega_R^{0}$ and obtained from the photonic spectral weight of the hybrid eigenstates. The spectra are normalized to the maximum of the disorder-free ($f=0$) spectrum; the same normalization factor is applied to all disorder strengths. The dashed curves show the Gaussian site-energy distributions of the corresponding uncoupled ensembles, normalized to unit maximum and scaled by a common factor of 0.45 for comparison. The logarithmic-scale inset shows the same spectra in the central molecular region, where disorder transfers weak photonic weight to states near $\omega_M$ that is not apparent on the linear scale. Energies are plotted relative to $\omega_M$ and reported in cm$^{-1}$. Simulation parameters are $N=1000$, $\omega_C=\omega_M$, and $\Omega_R^{0}=200~\mathrm{cm}^{-1}$. Further numerical details are given in SI Sec.~S1. The dependence of the plotted spectra on the energy-bin width is examined in \si{SI Fig.~S1}, while the disorder dependence of $\widetilde{\Omega}_R$ and of the photonic weight near $\omega_M$, including their sensitivity to the form of the disorder distribution, is shown in \si{SI Fig.~S2.}}
	\label{fig:fig1}
\end{figure}

We model the cavity-coupled molecular ensemble using a disordered Tavis--Cummings Hamiltonian within the rotating-wave approximation. The model describes an ensemble of $N$ non-interacting two-level molecules collectively coupled to a single quantized cavity mode\cite{Tavis1968}. Each molecule is represented by a ground state $\ket{{\rm g}_j}$ and an excited state $\ket{{\rm e}_j}$, with transition energy $\omega_j$, where $j=1,\ldots,N$. The cavity is treated as a single bosonic mode with resonance energy $\omega_C$, and the arrangement is shown schematically in the left panel of Fig.~\ref{fig:fig1}(a). We set $\hbar=1$ throughout, so that frequencies and energies are used interchangeably, and report all numerical values as wavenumbers in cm$^{-1}$.
 
We focus on the linear optical regime, where the excitation density is sufficiently low that the relevant dynamics are restricted to the zero- and one-excitation manifold. This approximation is appropriate for weak optical probing and reduces the coupled light--matter problem to an $(N+1)$-dimensional single-excitation Hilbert space. The collective ground state of the coupled cavity--molecule system is $\ket{G} = (\prod_{j=1}^{N}\ket{{\rm g}_j})\otimes \ket{0}$, where all molecules are in their electronic ground state and $\ket{0}$ denotes the zero-photon (vacuum) state of the cavity. The single-excitation basis consists of the one-photon cavity state $\ket{L} = (\prod_{j=1}^{N}\ket{{\rm g}_j})\otimes \ket{1}$, together with the molecular single-excitation states $\ket{M_j} = \ket{{\rm e}_j}(\prod_{i\neq j}^{N}\ket{{\rm g}_i})\otimes \ket{0}$, in which only the $j$th molecule is electronically excited and the remaining $N-1$ molecules are in their ground state. In the basis $\{\ket{L},\ket{M_1},...,\ket{M_N}\}$, the coupled light--matter Hermitian Hamiltonian is
\begin{equation}
	\hat{H} = \omega_C\ket{L}\bra{L} + \sum_{j=1}^{N} \omega_j \ket{M_j}\bra{M_j}
	+ \sum_{j=1}^{N} g_j \left(\ket{L}\bra{M_j} + \ket{M_j}\bra{L} \right),
	\label{eq:1}
\end{equation}
where $g_j$ denotes the light--matter coupling strength of molecule $j$. In the calculations below, the coupling is taken to be identical for all molecules, $g_j=g_c$. This Hamiltonian describes coherent exchange of excitation between the cavity photon and molecular excitations. The uniform coupling approximation neglects variations in $g_j$ arising, for example, from orientational disorder of the molecular transition dipoles or from the spatial profile of the cavity field\cite{Cwik2016,HerreraSpano2018}. Fixing $g_j=g_c$ allows us to isolate the effect of energetic disorder without introducing an additional source of inhomogeneity. 

To incorporate static energetic disorder, the molecular transition energies are assumed to follow a Gaussian distribution $P(\omega_j) = \frac{1}{\sqrt{2\pi\sigma^2}} \exp\!\left[-\frac{(\omega_j-\omega_M)^2}{2\sigma^2}\right],$ where $\omega_M$ is the mean molecular transition energy and $\sigma$ is the standard deviation of the site-energy distribution. Thus, $\sigma$ describes static site-to-site energetic disorder rather than homogeneous dephasing or dynamical fluctuations. A Gaussian distribution provides a standard description of inhomogeneous broadening arising from many weak environmental contributions to the molecular transition energies\cite{Knapp1984,Bassler1993,Bondarenko2020,Lev2022}. Its unbounded tails are, however, relevant to the localization problem considered here, because they provide molecular transitions progressively closer to the polariton energies as the disorder increases. A bounded distribution need not provide the same near-resonant molecular states at an equivalent width. We therefore do not regard the numerical crossover obtained for Gaussian disorder as universal, and in Sec.~II.D compare it with bounded uniform distributions matched to the Gaussian in either variance or full width at half maximum (FWHM). For the energy-ordered correlation analysis introduced in Sec.~II.D, the sampled molecular transition energies are sorted within each disorder realization, so that the molecular index $j$ labels position within the disorder broadened energy distribution rather than physical position. This relabeling has no effect on the Hamiltonian spectrum or the ordering-independent localization measures introduced below.

In the disorder-free limit, $\sigma=0$, all molecules are energetically identical. The cavity then couples only to the permutation-symmetric collective molecular excitation, $\ket{B} = \frac{1}{\sqrt{N}}\sum_{j=1}^{N}\ket{M_j}$, commonly referred to as the bright state. Because all molecular transition dipoles add coherently in this state, the effective light--matter coupling is enhanced from $g_c$ to $g_c\sqrt{N}$. Diagonalization of Eq.~\ref{eq:1} in the symmetric subspace spanned by $\ket{L}$ and $\ket{B}$ gives two hybrid light--matter eigenstates, the lower polariton (LP) and upper polariton (UP),
\begin{equation}
	\ket{\psi_{\text{LP}}}\equiv \ket{-} = \cos(\theta)\ket{B} - \sin(\theta)\ket{L}, \\
	\ket{\psi_{\text{UP}}}\equiv \ket{+} = \sin(\theta)\ket{B} + \cos(\theta)\ket{L}
	\label{eq:2}
\end{equation}
where the mixing angle satisfies
$\theta=\frac{1}{2}\tan^{-1}\!\left[2g_c\sqrt{N}/(\omega_C-\omega_M)\right]$,
with $\theta=\pi/4$ at zero detuning. The corresponding LP--UP energy separation is $\omega_{\text{UP}}-\omega_{\text{LP}} = \sqrt{(\omega_C-\omega_M)^2 + 4g_c^2N}$. At zero detuning, $\omega_C=\omega_M$, this separation reduces to the \emph{bare collective coupling} $\Omega_R^{0}=2\sqrt{\sum_{j=1}^{N}|g_j|^2}=2g_c\sqrt{N}$, which is held fixed throughout the calculations. We emphasize that $\Omega_R^{0}$ is a parameter of the underlying Hamiltonian and is distinct from the disorder-dependent separation of the spectral maxima, denoted $\widetilde{\Omega}_R$, which is discussed in Sec.~II.C. Accordingly, we characterize the energetic disorder by the normalized quantity $f=\sigma/\Omega_R^{0}$. Small $f$ corresponds to a regime in which collective hybridization dominates over energetic disorder, whereas larger $f$ indicates that the inhomogeneous molecular broadening becomes increasingly comparable to the collective light--matter coupling scale.

The remaining $N-1$ molecular superposition states are orthogonal to $|B\rangle$ and do not couple directly to the cavity photon in the homogeneous limit. These states form a dark-state manifold, denoted as $\ket{D_\alpha}$ with $\alpha=1,\ldots,N-1$, and are degenerate at $\omega_M$ when $\sigma=0$. This eigenstate structure is illustrated schematically in Fig.~\ref{fig:fig1}(a). Static energetic disorder breaks the permutation symmetry of the molecular ensemble, lifts the dark-state degeneracy, and mixes the bright and dark molecular subspaces. Consequently, oscillator strength is redistributed among a larger number of hybrid eigenstates, including weakly bright dark-state-derived states, while the observed spectral separation $\widetilde{\Omega}_R$ can differ from the bare collective coupling $\Omega_R^{0}$. These spectral consequences are examined in Sec.~II.C.
\FloatBarrier


Unless otherwise stated, all calculations use $N=1000$ molecules at zero detuning, with the bare collective coupling fixed at $\Omega_R^{0}=200~\mathrm{cm}^{-1}$, corresponding to $g_c\simeq3.16~\mathrm{cm}^{-1}$, and are averaged over 100 independent disorder realizations per point. When $N$ is varied, the single-molecule coupling is scaled as $g_c\propto N^{-1/2}$ so that $\Omega_R^{0}$ remains unchanged. Further numerical details, convergence tests, sensitivity to the analysis protocol, and the bounded-distribution calculations are provided in the Supporting Information. Before presenting the spectral and localization results, we briefly clarify the scope and limitations of the effective model used here.

\subsection{Scope and Limitations of the Model}

The disordered Tavis--Cummings Hamiltonian of Eq.~\ref{eq:1} provides an effective description in which collective light--matter coupling, energetic disorder, and, later, finite linewidths can be considered on the same footing. The molecular excitations are represented as non-interacting two-level emitters coupled through a common cavity mode. Direct intermolecular excitonic coupling and short-range electronic interactions are therefore not included explicitly. In molecular aggregates these interactions can themselves produce exciton delocalization, H- and J-type spectral shifts, and exchange-narrowing effects\cite{Knapp1984,Spano2010,HestandSpano2018}. The present description is consequently most directly applicable when such intermolecular interactions are weak, or when their average effects can be absorbed into effective molecular transition energies and couplings. In dense molecular films and liquids, additional effects associated with intermolecular interactions, local dielectric environment, molecular orientation, and concentration-dependent optical properties can modify both the bare molecular response and its coupling to the cavity\cite{Cwik2016,HerreraSpano2018,Arkajit2023Rev}. Recent experiments under vibrational and electronic strong coupling have also reported changes in observables sensitive to intermolecular dispersion and ground-state intermolecular interactions\cite{Ebbesen2024VSCinNMR,Biswas2025ESC}. Such effects involve microscopic intermolecular and electronic-structure physics that is not contained explicitly in the effective two-level model used here. These effects are not treated explicitly here, since our aim is to isolate the consequences of energetic disorder within the collective coupling problem.

The effective two-level description used in our model does not retain the full antisymmetrized many electron structure of the constituent molecules. It therefore cannot describe microscopic changes of electronic structure arising from orbital overlap, Pauli exchange, or cavity-induced redistribution of the local electronic polarization. More microscopic cavity quantum electrodynamics (QED) approaches, including \textit{ab initio} Pauli--Fierz and QED-based electronic structure treatments, have shown that collective strong coupling can modify local molecular polarization and related electronic properties even in the many molecule limit\cite{Sidler2021JPCL,Sidler2024JPCL,Horak2025PRR,Castagnola2025PRX,KochRubioJCP2021AbInitioStudy}. Such effects are complementary to the present question. Here the molecular basis states are taken as given, and we ask how the molecular component of a hybrid excitation is distributed among the members of a disordered ensemble. The localization measures introduced below should therefore be interpreted as properties of these effective molecular excitations rather than as measures of cavity-induced changes in the internal electronic structure of an individual molecule.

The calculations are further restricted to the weak excitation, linear-response regime. Within the rotating-wave Tavis--Cummings model the ground state remains the uncoupled molecular--cavity vacuum state, and the present treatment therefore does not address cavity-induced modifications of ground state potential energy surfaces or ground state chemical reactivity\cite{Ribeiro2018,Feist2018,LiNitzan2022,Arkajit2023Rev}. Our results are most directly applicable to electronic strong coupling, where the polariton states are accessed optically from the ground state and the single-excitation manifold provides the natural description of the linear spectrum. For the electronic strong-coupling regime considered here, the negligible equilibrium thermal occupation of optical-frequency cavity photons at room temperature does not restrict the present analysis, because the single-excitation polariton states are created by the weak optical probe rather than assumed to be thermally populated. Formally, a similar eigenstate analysis can be constructed for vibrational strong coupling, but its chemical interpretation requires additional care: molecular anharmonicity, vibrational relaxation, intramolecular mode coupling, and the coupling of vibrations to a thermal environment are absent from the present model\cite{HerreraSpano2018,Ribeiro2018,Arkajit2023Rev}. We therefore do not extend the eigenstate delocalization criteria obtained below to vibrationally mediated ground state chemistry.

The energetic disorder considered in our model is static. Temperature is not an explicit variable of the Hamiltonian, and the disorder width $\sigma$ should not be identified directly with temperature. Experimentally, temperature can modify molecular conformations, solvent and orientational dynamics, homogeneous linewidths, and the balance between static and dynamic disorder. In particular, static and dynamical fluctuations can influence the polaritonic and dark-state manifolds in qualitatively different ways\cite{Catuto2025JCP,Qiang2026Arxiv}. Recent work has also shown that room temperature thermal disorder can qualitatively modify the photochemical consequences of electronic strong coupling\cite{Dutta2024ThermalDisorder}. Within the present model these environmental effects enter only phenomenologically through the static distribution of transition energies and, in Sec.~II.E, through the molecular and cavity linewidths. A microscopic finite-temperature description would require explicit dynamical disorder, vibrational degrees of freedom, or coupling to a thermal bath.

Within these boundaries, the model provides a controlled setting for the question addressed here: whether an optically visible polariton eigenstate retains a molecular component that is extended across the ensemble as energetic disorder and dissipation are varied. We first examine how static disorder modifies the cavity-probed spectrum.

\subsection{Linear Spectra under Static Energetic Disorder}

The effect of static energetic disorder on the polariton spectrum is shown in Fig.~\ref{fig:fig1}(b). We consider the linear, weak excitation regime and characterize the cavity-accessible spectral response through the photonic component of the hybrid eigenstates. For each disorder realization, we diagonalize the single-excitation Hamiltonian in Eq.~\ref{eq:1} and obtain eigenstates $\ket{\psi_n}$ with eigenenergies $E_n$. The photonic spectral weight of each eigenstate is calculated from its squared projection onto the cavity basis state $\ket{L}$, $p_n=|\langle L|\psi_n\rangle|^2$, which determines its visibility in a cavity-probed linear spectrum. The disorder-averaged spectral visibility is then
\begin{equation}
	I(\omega)=\left\langle \sum_{n=1}^{N+1} p_n\,\delta(\omega-E_n) \right\rangle_{\mathrm{dis}},
	\label{eq:3}
\end{equation}
where $\langle\cdots\rangle_{\mathrm{dis}}$ denotes averaging over independent static disorder realizations. Because Eq.~\ref{eq:3} is a sum of discrete eigenstate contributions, a finite energy bin is used only to display the disorder-averaged spectral envelope in Fig.~\ref{fig:fig1}(b), rather than imposing an artificial homogeneous linewidth. The absolute peak heights therefore depend on the chosen bin width (\si{SI Sec.~S1, Fig.~S1, and Table~S1}) and are not used as quantitative observables; instead, we focus on the peak positions and integrated photonic spectral weights.

Note that the quantity in Eq.~\ref{eq:3} represents the cavity spectral function and is not directly identical to an experimentally measured transmission or absorption spectrum. For a Fabry--P\'erot cavity, the measured transmission spectrum is related to the cavity response $I(\omega)$ through the mirror transmission and the frequency-dependent incoupling and outcoupling of the cavity mode\cite{GardinerCollett1985,Carmichael1993}. An experimentally measured absorption spectrum can additionally contain a direct molecular absorption contribution. Nevertheless, provided these external factors vary smoothly over the spectral range of interest, they do not significantly affect the polariton peak positions or the integrated spectral weights considered here. We therefore use these quantities to characterize the disorder-dependent spectral response. Eigenstates with appreciable photonic weight $p_n$ contribute to the cavity-probed spectrum, whereas states that are purely molecular in the homogeneous limit remain dark unless disorder gives them finite photonic character.

As shown in Fig.~\ref{fig:fig1}(b), at zero disorder, $f=0$, the spectrum consists of two sharp polariton resonances located symmetrically about $\omega_M$ at zero detuning and separated by the bare collective coupling $\Omega_R^{0}$. In this limit all molecules have the same transition energy, so the cavity couples only to the permutation-symmetric bright molecular state, while the $N-1$ dark states carry no photonic weight. Consequently, only the LP and UP appear in the cavity-projected spectrum.

As the disorder increases, the LP and UP features develop broader spectral envelopes while remaining clearly resolved over the range shown in Fig.~\ref{fig:fig1}(b). This broadening originates primarily from disorder-induced mixing between the bright and dark molecular subspaces rather than from fluctuations of the ensemble averaged transition energy, which decrease as $\sigma/\sqrt{N}$. In the bright--dark basis, the diagonal site-energy disorder generates off-diagonal couplings between $\ket{B}$ and the dark states $\ket{D_\alpha}$, thereby redistributing the photonic weight that is concentrated in the two polariton states at $f=0$ over an increasing number of hybrid eigenstates. States derived predominantly from the dark manifold consequently acquire weak photonic character and become visible in the cavity-projected spectrum. This redistribution is seen most clearly in the logarithmic-scale inset of Fig.~\ref{fig:fig1}(b), where weak spectral weight progressively appears around the bare molecular transition $\omega_M$. Quantitatively, the photonic weight integrated over $|\omega-\omega_M|\leq\Omega_R^{0}/4$ increases from zero at $f=0$ to approximately 0.04, 0.07, and 0.13 at $f=0.10$, 0.15, and 0.25, respectively. Thus, even while the spectrum remains dominated by the LP and UP resonances, an increasing fraction of the cavity-accessible spectral weight is carried by disorder-mixed molecular states.

Further, the LP--UP peak separation increases moderately with disorder. At $f=0$ the exact separation is $\Omega_R^0$ by construction. From the finite disorder spectral maxima we obtain $\widetilde{\Omega}_R/\Omega_R^0\simeq1.02$, 1.05, and 1.14 at $f=0.10$, 0.15, and 0.25, respectively; the finite energy-bin
resolution is examined in SI Sec.~S1 and Table~S1. This increase should not be interpreted as an enhancement of the microscopic coupling strength $g_c$. Rather, static disorder couples the polariton branches to molecular states lying predominantly between them, producing level repulsion that shifts the UP to higher energy and the LP to lower energy. In the perturbative regime, the disorder-induced shift scales approximately as $\sigma^2/\Omega_R^{0}$ for each branch,\cite{Sebastian2022} giving a corresponding increase of order $2\sigma^2/\Omega_R^{0}$ in the LP--UP separation. The disorder-dependent spectral splitting $\widetilde{\Omega}_R$ can therefore exceed the bare collective coupling $\Omega_R^{0}$, further emphasizing that the two quantities should not be used interchangeably. Consequently, inferring the collective coupling directly from the observed peak separation can lead to an overestimate of $\Omega_R^{0}$ in a disordered ensemble.

Figure~\ref{fig:fig1}(b) therefore presents an important contrast. Static energetic disorder substantially redistributes photonic spectral weight and modifies the apparent polariton splitting, yet a clearly resolved LP--UP doublet persists throughout the disorder range shown. On the basis of the spectrum alone, the system would therefore continue to appear strongly coupled. The spectrum, however, does not determine how broadly the molecular component of those cavity-visible eigenstates is distributed across the ensemble. We address this question next at the level of the molecular eigenvectors.

\subsection{Disorder-Induced Localization of Polaritons}
To quantify the molecular extent of these cavity-visible states, we analyze the molecular component of each hybrid eigenstate. We write the $n$th eigenstate as
\begin{equation}
	\ket{\psi_n}=\ell_n\ket{L}+\sum_{j=1}^{N} a_{j}^{(n)}\ket{M_j},
	\label{eq:4}
\end{equation}
where $\ell_n$ is the photonic amplitude and $p_n=|\ell_n|^2$ is the corresponding photonic weight. Since the localization considered here refers specifically to the molecular component, the molecular amplitudes are first renormalized within the molecular subspace, $\tilde{a}_{j}^{(n)}=a_{j}^{(n)}/\sqrt{\sum_{k}|a_k^{(n)}|^2}$. This removes the overall excitonic Hopfield weight and isolates the distribution of molecular amplitude within the ensemble. From these normalized amplitudes we construct the molecular exciton density matrix of the state,
\begin{equation}
	\rho_{ij}^{(n)}=\tilde{a}_{i}^{(n)}\tilde{a}_{j}^{(n)*} ,
	\label{eq:5}
\end{equation}
whose diagonal elements describe the molecular population distribution and whose off-diagonal elements describe coherence between molecular excitation amplitudes. We use three complementary measures constructed from this molecular component, each of which weights the eigenvector differently, as discussed below.

\subsubsection{Participation ratio}
A widely used measure of wavefunction localization in disordered quantum systems is the participation ratio (PR), or equivalently the inverse participation ratio. The PR estimates the effective number of basis states carrying significant population and has been widely used in studies of Anderson localization and exciton localization in molecular systems\cite{Anderson1958,Murphy2011,KramerMacKinnon1993,Bondarenko2020}. For the normalized molecular component of eigenstate $n$, the PR is
\begin{equation}
	{\rm PR}^{(n)}
	=
	\frac{1}
	{\displaystyle\sum_{j=1}^{N}
		\left|\widetilde{a}_j^{(n)}\right|^4}
	=
	\frac{1}
	{\displaystyle\sum_{j=1}^{N}
		\left(\rho_{jj}^{(n)}\right)^2}.
	\label{eq:6}
\end{equation}
It equals unity for an excitation localized on a single molecule and $N$ for a state uniformly distributed over the entire ensemble. More generally, a state of equal amplitude over $M$ molecules gives ${\rm PR}=M$. Thus, ${\rm PR}/N$ provides a direct measure of the fraction of the molecular ensemble participating significantly in the eigenstate. In the homogeneous limit, the molecular component of both polariton branches is the symmetric bright state $\ket{B}$, giving ${\rm PR}/N=1$.

\subsubsection{Molecular coherence measure}
The PR is constructed only from the diagonal populations of the molecular density matrix and is therefore insensitive to its off-diagonal elements. These off-diagonal elements characterize correlations between molecular excitation amplitudes and are closely related to excitonic coherence and collective optical response in molecular systems\cite{Meier1997,Cho2008ChemRev,Scholes2006,Jiang2023}. To complement the population-based PR, we therefore define the molecular coherence measure
\begin{equation}
	L_{\rho}^{(n)}
	=
	\sum_{i,j=1}^{N}
	\left|\rho_{ij}^{(n)}\right|
	=
	\left(
	\sum_{j=1}^{N}
	\left|\widetilde{a}_j^{(n)}\right|
	\right)^2 .
	\label{eq:7}
\end{equation}
Because the absolute values of the density-matrix elements are used, $L_\rho$ quantifies the extent of the molecular coherence support without retaining information about the relative phases of the amplitudes. For a molecular state uniformly distributed with equal amplitude over $M$ molecules, $L_\rho=M$, while a state localized on a single molecule gives $L_\rho=1$. Thus, $L_\rho$ is calibrated on the same effective molecular extent scale as the PR for this benchmark, and
$L_\rho/N$ gives the corresponding fractional extent. The common calibration of PR, $L_\rho$, and the correlation-width measure introduced below is demonstrated in \si{SI Sec.~S2 and Fig.~S3}.

Although PR and $L_\rho$ share the same limiting values, they weight a nonuniform molecular eigenvector differently. This distinction follows directly from the molecular amplitude structure of a hybrid eigenstate. Projecting the eigenvalue equation associated with Eq.~\ref{eq:1} onto the $j$th molecular state gives,
\begin{equation}
	a_j^{(n)}
	=
	\frac{g_c\ell_n}{E_n-\omega_j},
	\label{eq:8}
\end{equation}
for an eigenstate with finite photonic amplitude. A detailed derivation and representative molecular amplitude profiles are given in \si{SI Sec.~S3 and Fig.~S4}. Equation~\ref{eq:8} shows that a molecular transition lying close to the eigenenergy can acquire a much larger amplitude than more distant transitions, while weaker amplitudes remain distributed over the rest of the ensemble. Because the PR
contains $|\widetilde a_j|^4$, it is weighted strongly toward the largest molecular populations, whereas $L_\rho$ depends on the summed amplitude magnitude and therefore remains more sensitive to weaker amplitudes distributed across the ensemble. The two measures thus provide complementary views of the molecular extent of a nonuniform polariton eigenstate.

\subsubsection{Correlation width in the energy-ordered molecular basis}
Taken together, ${\rm PR}$ and $L_\rho$ provide complementary measures of the molecular extent of a polariton eigenstate, with the former emphasizing the population distribution and the latter being more sensitive to weaker molecular amplitudes across the ensemble. However, neither measure resolves how these molecular amplitudes remain correlated across the disorder-broadened distribution of transition energies. To visualize this structure, we calculate an autocorrelation function in the energy-ordered molecular basis, analogous to those used to characterize wavefunction correlations in disordered molecular aggregates\cite{Bondarenko2020,Kuhn1997,Pullerits2001}. For each normalized molecular eigenvector we define
\begin{equation}
C_n(\Delta j) = \left|\sum_j\widetilde a_j^{(n)}\widetilde a_{j+\Delta j}^{(n)*}\right|,
	\label{eq:9}
\end{equation}
with $C_n(0)=1$. We define the corresponding correlation width $\xi_n$ as the discrete FWHM of $C_n(\Delta j)$.

The interpretation of $\xi$ differs from that of a conventional spatial localization length. The Hamiltonian in Eq.~\ref{eq:1} contains no molecular positions or intermolecular hopping, and the index $j$ is assigned after sorting the molecular transition energies in ascending order. Thus, $\Delta j$ measures separation along the energy-ordered molecular basis rather than physical distance in real space. We therefore refer to $\xi$ as a \emph{spectral-index correlation width}. A large $\xi$ indicates that the molecular amplitudes remain correlated across a broad portion of the ordered transition-energy distribution, whereas a small $\xi$ indicates that appreciable correlations are confined to a much narrower subset of that distribution.

Because a fixed interval in the energy-ordered index need not correspond to the same physical energy interval as the ensemble size or local density of molecular states changes, we additionally construct an ordering-independent correlation directly as a function of a prescribed molecular transition-energy separation $\Delta\omega$. The resulting normalized energy-space width closely tracks the energy-ordered index result (\si{SI Sec.~S9 and Fig.~S11}). Moreover, PR and $L_\rho$ are independent of any molecular ordering, so the central localization conclusions do not rely on the energy sorting used for Eq.~\ref{eq:9}.

\subsubsection{Optically visible eigenstates and the photon-weighted bright sector}

The measures introduced above characterize the molecular component of an individual eigenstate. Under energetic disorder, however, the photonic weight associated with the homogeneous bright state is redistributed among several polariton-like and weakly bright eigenstates. The molecular extent of an individual optically visible eigenstate therefore need not be identical with the extent obtained when all states carrying appreciable photonic weight are considered collectively. We consequently analyze the molecular delocalization at both the \emph{individual-eigenstate level} and at the level of the photon-weighted bright sector. Here, the term \emph{bright sector} refers to the set of disorder-mixed eigenstates that carry appreciable photonic weight, and should not be confused with the single symmetric bright state $\ket{B}$ of the homogeneous limit.

At the eigenstate level, we retain states according to their photonic visibility, $p_n \geq \eta p_{\rm max}$, within each disorder realization, with $\eta=0.1$. The localization measure is evaluated for each retained state and then averaged over the selected states and disorder realizations. This criterion selects the optically active polariton-derived states while excluding the large number of nearly dark eigenstates with negligible cavity contribution. Its sensitivity to the choice of $\eta$ is examined over a broad range in \si{SI Fig.~S5}.

To characterize the same optically accessible states collectively, we construct a photon-weighted molecular density matrix. Defining the normalized molecular component of eigenstate $n$ as $\ket{\phi_n} = \sum_{j=1}^{N} \widetilde{a}_j^{(n)}\ket{M_j}$, the corresponding bright sector density matrix is
\begin{equation}
	\rho_{\rm B}
	=
	\frac{
		\displaystyle\sum_{n\in\mathcal{S}}
		p_n\,\ket{\phi_n}\bra{\phi_n}
	}{
		\displaystyle\sum_{n\in\mathcal{S}}p_n
	},
	\label{eq:10}
\end{equation}
where $\mathcal{S}$ denotes the set of states satisfying the same photonic visibility criterion, $p_n\geq\eta p_{\rm max}$. The PR, molecular coherence measure $L_\rho$, and spectral-index correlation width are then evaluated from $\rho_{\rm B}$ using the corresponding density matrix definitions, providing a collective measure of the molecular extent of the photon-weighted bright sector. As a threshold-independent check, we repeat the construction using fixed spectral windows around the LP and UP, including all eigenstates within those windows irrespective of their individual photonic weight (\si{SI Sec.~S8 and Fig.~S10}).

\subsubsection{Disorder dependence of molecular delocalization}

\begin{figure}[ht!]
	\centering
	\includegraphics[width=\textwidth]{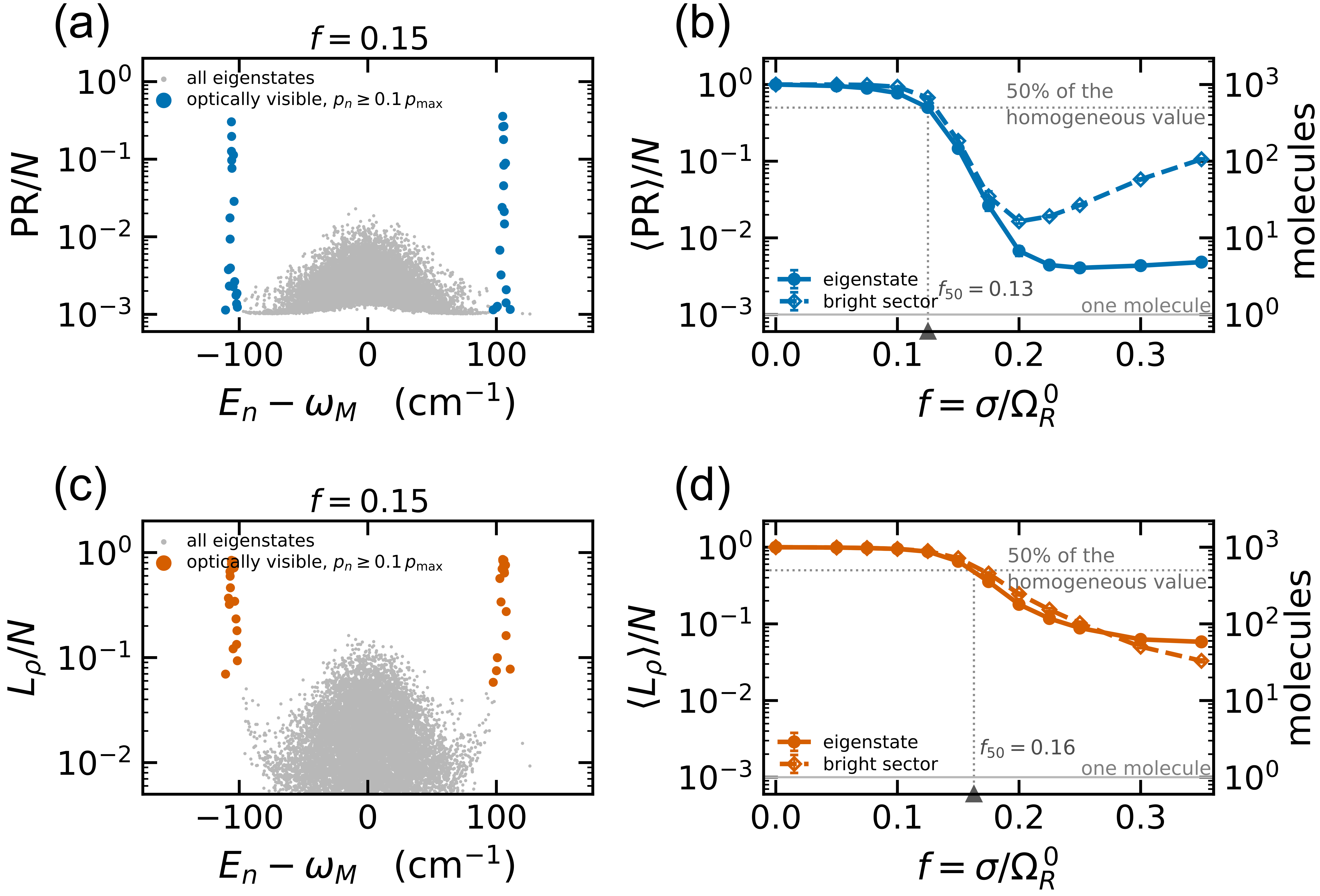}
	\caption{\footnotesize Molecular extent of the optically visible polariton-derived states under static energetic disorder. (a) Normalized molecular participation ratio, ${\rm PR}/N$, of the hybrid eigenstates at $f=0.15$. For clarity, 12 independent disorder realizations are displayed without averaging. Gray points show all $N+1$ eigenstates from each realization, while colored points identify the subset satisfying the photonic-visibility criterion $p_n\geq0.1\,p_{\rm max}$. (b) Disorder-averaged ${\rm PR}/N$ of the optically visible states as a function of disorder strength. Filled symbols show the average obtained by evaluating the PR for each selected eigenstate separately, while open symbols show the corresponding photon-weighted bright-sector result. The dotted line denotes 50\% of the homogeneous-limit value and the triangle marks the corresponding crossover disorder $f_{50}$. The right axis expresses the same measure as an effective number of participating molecules. (c) Corresponding molecular coherence measure, $L_\rho/N$, for the same eigenstates and disorder realizations as in (a). (d) Corresponding disorder-averaged $L_\rho/N$ of the optically visible states. For a molecular state of equal amplitude over $M$ molecules, both PR and $L_\rho$ equal $M$ (\si{SI Fig.~S3}). The results in (b) and (d) are averaged over 100 independent disorder realizations; error bars denote standard errors of the mean and are smaller than the symbols where not visible. \si{Dependence on the photonic-visibility threshold and convergence with the number of disorder realizations are shown in SI Figs.~S5 and S6, respectively; comparison with a fixed LP/UP spectral-window construction is given in SI Fig.~S10.}}
	\label{fig:fig2}
\end{figure}

Having defined the localization measures and the states over which they are evaluated, we now return to the central question raised by the spectra in Fig.~\ref{fig:fig1}(b): how broadly distributed is the molecular component of the cavity-visible states as energetic disorder increases? Although the LP and UP remain clearly resolved over the disorder range considered above, their spectral visibility does not establish that their molecular components remain extended over the ensemble. Figure~\ref{fig:fig2} therefore examines the same disorder evolution directly at the level of the molecular eigenvectors, using the PR and $L_\rho$ as complementary measures of their extent.

\FloatBarrier
Figures~\ref{fig:fig2}(a) and (c) show the PR and $L_\rho$, respectively, of the hybrid eigenstates at $f=0.15$ for 12 disorder realizations, displayed
without averaging to illustrate the state-to-state distribution of molecular
extent. The states carrying appreciable photonic weight occur predominantly near the polariton branches and generally have a larger molecular extent than the predominantly dark states. Nevertheless, even within this cavity-visible subset the molecular extent varies substantially: some states remain distributed over a large fraction of the ensemble, whereas others with appreciable photonic weight are concentrated on a much smaller subset of molecules. Thus, photonic visibility does not directly determine how many molecules participate in a given hybrid eigenstate.

The evolution with disorder is shown in Figs.~\ref{fig:fig2}(b) and (d), where the PR and $L_\rho$ are averaged over the optically visible eigenstates and normalized by $N$. At weak disorder both measures remain close to unity, as expected for molecular components that retain the character of the homogeneous symmetric bright state. With increasing $f$, however, both decrease rapidly, indicating a crossover from an extended collective molecular component to one concentrated over a progressively smaller fraction of the ensemble.

To characterize the crossover, we define $f_p$ as the disorder strength at which the normalized molecular extent measure reaches the level $p/100$. These levels correspond approximately to retaining $p\%$ of the homogeneous molecular extent on the normalization used for the three measures. The dotted lines and triangular markers in Figs.~\ref{fig:fig2}(b) and (d) show the half-value level $f_{50}$. For $N=1000$, the present calculation gives $f_{50}\simeq0.13$ for the PR and $f_{50}\simeq0.16$ for $L_\rho$. Because both measures give the value $M$ for a state uniformly distributed over $M$ molecules, the 50\% level corresponds to an effective molecular extent of approximately 500 molecules for $N=1000$. The quantity $f_{50}$ should therefore be viewed as a convenient midpoint of the finite-size crossover rather than as a sharp localization threshold; lower fractional levels such as $f_{20}$ and $f_{10}$ characterize later stages of the same crossover.

The somewhat larger $f_{50}$ obtained from $L_\rho$ reflects the different weighting of the molecular eigenvector by the two measures. The PR is dominated by the highly populated core of the molecular component, whereas $L_\rho$ remains sensitive to weaker amplitudes distributed over a broader part of the ensemble. Consequently, $L_\rho$ decreases more gradually with disorder. This distinction becomes particularly clear at stronger disorder: at $f=0.25$, the optically visible eigenstates have an average PR of about four molecules, whereas $L_\rho$ remains about 90 molecules. These values do not represent conflicting localization scales, but rather the different sensitivity of the two measures to the compact population core and the weaker molecular amplitude tails discussed above and in \si{SI Sec.~S3}.

The photon-weighted bright sector results, shown by the open symbols in Figs.~\ref{fig:fig2}(b) and (d), follow the same decrease through the crossover and yield essentially the same crossover scale as the individual eigenstate averages. The loss of molecular extent is therefore not a consequence of evaluating the visible eigenstates separately. At larger disorder the bright-sector PR can turn upward because population from many individually localized states is combined in the mixed density matrix, without a corresponding recovery of the off-diagonal molecular correlations. This behavior and the fixed LP/UP-window construction are examined in \si{SI Sec.~S8 and Fig.~S10}.

\subsubsection{Crossover scale and its robustness}
Having established the disorder-induced reduction of molecular extent from the PR and $L_\rho$, we next characterize the same crossover through correlations across the disorder-broadened molecular energy distribution and examine its dependence on system size and on the form of the energetic disorder.

\begin{figure}[ht!]
	\centering
	\includegraphics[width=\textwidth]{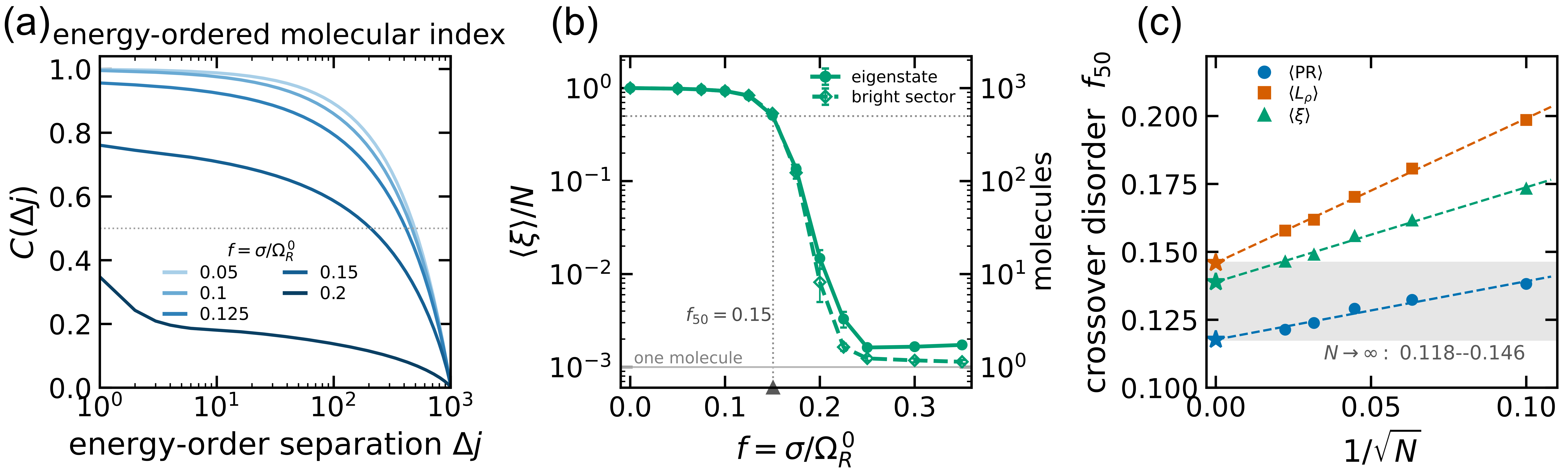}
	\caption{\footnotesize Correlation based characterization and finite-size scaling of the molecular delocalization crossover. (a) Normalized molecular correlation function $C(\Delta j)$ for representative disorder strengths, averaged over the optically visible polariton states. The molecular states within each realization are ordered by transition energy, so $\Delta j$ denotes separation along this energy ordering rather than physical distance. The discrete FWHM defines the spectral-index correlation width $\xi$. An ordering-independent correlation constructed directly from molecular transition-energy separations is compared in \si{SI Sec.~S9 and Fig.~S11}. (b) Disorder dependence of the normalized correlation width, $\langle\xi\rangle/N$, evaluated at the individual eigenstate and photon-weighted bright sector levels. The dotted horizontal line denotes the normalized level $\langle\xi\rangle/N=0.5$ and the triangle marks the corresponding crossover disorder $f_{50}$. Error bars are standard errors of the mean over 100 disorder realizations and are smaller than the symbols where not visible. (c) Finite-size dependence of $f_{50}$ obtained from PR, $L_\rho$, and $\xi$ for $N=100$--2000, with $\Omega_R^0$ held fixed by scaling $g_c\propto N^{-1/2}$. Linear extrapolations in $1/\sqrt{N}$ are shown, with stars marking the corresponding $N\to\infty$ intercepts. The full finite-size curves, crossover widths, and sensitivity to the extrapolation form are given in \si{SI Sec.~S6, Figs.~S7--S8, and Table~S4}.}
	\label{fig:fig3}
\end{figure}

Figure~\ref{fig:fig3}(a) shows the normalized correlation function $C(\Delta j)$ for representative disorder strengths. At weak disorder the correlations extend over a broad range of the energy-ordered molecular basis, consistent with a molecular component distributed across much of the ensemble. Increasing $f$ progressively narrows this profile, indicating that appreciable correlations become confined to molecules lying within a much smaller region of the ordered transition energy distribution. As emphasized above, $\Delta j$ denotes separation in this energy-ordered basis rather than a real space distance; the corresponding width $\xi$ is therefore a spectral-index correlation width rather than a spatial localization length.

The change of $\xi/N$ with disorder is shown in Fig.~\ref{fig:fig3}(b). Like the PR and $L_\rho$, the normalized correlation width remains unity at weak disorder and then decreases rapidly as the static energetic disorder is increased. Applying the same fractional criterion introduced above gives $f_{50}\simeq0.15$ for $\xi$ at $N=1000$, compared with approximately $0.13$ and $0.16$ for the PR and $L_\rho$, respectively. Thus, although the three measures weight the molecular eigenvector differently, they place the half-value level of the loss of molecular extent within the same relatively narrow disorder range. The modest measure-dependent spread in $f_{50}$ is consistent with a smooth crossover rather than a sharply defined localization transition.

\FloatBarrier 

To determine whether this characteristic disorder scale  depends on the finite ensemble size as considered above, we repeat the calculations for $N=100$, $250$, $500$, $1000$, and $2000$, while keeping the bare collective coupling $\Omega_R^0$ fixed by scaling $g_c\propto N^{-1/2}$. Figure~\ref{fig:fig3}(c) shows the resulting $f_{50}$ values for all three measures. Although the crossover position shifts somewhat for the smaller ensembles, the variation becomes progressively weak as $N$ increases. Extrapolation against $N^{-1/2}$ gives large-$N$ \as{half-value positions of approximately $0.12$, $0.15$, and $0.14$ for the PR, $L_\rho$, and $\xi$, respectively}. Alternative extrapolation variables broaden the overall large-$N$ range to approximately $f=0.11$--0.16 (\si{SI Sec.~S6, Fig.~S8, and Table~S4}). 

For the Gaussian disorder considered here, these large-$N$ values imply that preserving approximately 50\% of the homogeneous molecular extent requires a bare collective coupling $\Omega_R^0/\sigma$ of order $7$. This requirement is considerably more stringent than that suggested by the persistence of a spectroscopically resolved LP--UP doublet, showing that substantial eigenstate-level molecular delocalization generally requires stronger collective coupling than spectral strong coupling signatures alone would indicate. The corresponding $f_{20}$ and $f_{10}$ values characterize progressively later stages of the same crossover and exhibit the same overall finite-size behavior; their full $N$ dependence is given in \si{SI Figs.~S7 and S8}.

The precise crossover scale, however, depends on the form of the energetic disorder distribution. This dependence can be understood directly from the molecular amplitude relation given in Eq.~\ref{eq:8}; molecular transitions lying close to a polariton eigenenergy acquire disproportionately large amplitudes and can therefore concentrate the molecular component onto a relatively small subset of the ensemble. The unbounded Gaussian tails increase the availability of molecular transitions close to the polariton eigenenergies as the disorder grows, whereas truncating the distribution reduces this near-resonant spectral weight and shifts the localization crossover to larger disorder. Consistent with this picture, bounded uniform distributions matched either in variance or in FWHM produce substantially larger crossover values (\si{SI Sec.~S7, Fig.~S9, and Table~S5}). The quantitative crossover scale is therefore distribution dependent; however, the central conclusion is robust across the distributions considered here: a well-resolved polariton spectrum can persist even when the molecular components of the cavity-visible eigenstates have lost a substantial fraction of their ensemble-wide delocalization.\\

Taken together, Figs.~\ref{fig:fig1}--\ref{fig:fig3} establish a clear separation between spectroscopic strong coupling and eigenstate-level molecular delocalization in the Hermitian disordered ensemble. The LP--UP doublet remains clearly visible, and its apparent spectral separation can even increase with disorder, while the PR, $L_\rho$, and spectral-index correlation width all show a pronounced loss of molecular extent. The precise crossover scale depends on the localization measure and on the statistics of energetic disorder, but the separation between spectral visibility and molecular delocalization is robust.


\subsection{Non-Hermitian Extension: Finite Cavity and Molecular Linewidths}

Real cavity--molecule systems are open: the cavity field decays through photon leakage, while molecular excitations possess finite homogeneous linewidths arising from relaxation, dephasing, and other dissipative processes\cite{Mukamel1995,MayKuhn2011,GardinerCollett1985,Carmichael1993}. Having established the separation between spectral visibility and molecular extent under static disorder, we now ask how finite and generally unequal cavity and molecular linewidths modify this relationship.

To incorporate these finite lifetimes phenomenologically within the linear response model, we replace the real bare resonance frequencies by complex energies,
\begin{equation}
	\omega_C\rightarrow\widetilde{\omega}_C=\omega_C-i\gamma_C,
	\qquad
	\omega_j\rightarrow\widetilde{\omega}_j=\omega_j-i\gamma_M ,
	\label{eq:NHfreq}
\end{equation}
where $\gamma_C$ and $\gamma_M$ denote the cavity and molecular homogeneous half-width-at-half-maximum (HWHM) linewidths, respectively. The corresponding full linewidths are therefore $2\gamma_C$ and $2\gamma_M$. Both imaginary contributions are taken to be negative, so the effective Hamiltonian describes passive decay without gain. This phenomenological non-Hermitian description is widely used to represent the complex resonance energies of open optical and polaritonic systems in the weak excitation regime\cite{Moiseyev2011,ElGanainy2018,Jonathan2022,Daniel2023Plexciton,Gerrit2024JCP}.

With Eq.~\ref{eq:NHfreq}, the disordered Hamiltonian in Eq.~\ref{eq:1} becomes
\begin{equation}
	\hat H_{\rm eff} = \widetilde{\omega}_C\ket{L}\bra{L} + \sum_{j=1}^{N} \widetilde{\omega}_j\ket{M_j}\bra{M_j}
	+ \sum_{j=1}^{N} g_c \left(\ket{L}\bra{M_j} + \ket{M_j}\bra{L} \right),
	\label{eq:Heff}
\end{equation}
whose eigenvalues are complex, $\widetilde E_n = E_n - i \Gamma_n$, where $E_n$ is the resonance energy and $\Gamma_n$ is the corresponding HWHM linewidth. For the localization analysis, the molecular components of the right eigenvectors are normalized over the molecular subspace and the same PR, $L_\rho$, and spectral-index correlation measures introduced above are evaluated.

It is useful to first consider the homogeneous limit, $\sigma=0$, which provides a simple analytical reference for the effect of finite linewidths. In this limit all molecules have the same complex transition energy $\widetilde{\omega}_M=\omega_M-i\gamma_M$. As in the Hermitian case, only the symmetric bright state $\ket{B}$ couples directly to the cavity, while the $N-1$ dark states remain uncoupled with complex energy $\widetilde{\omega}_M$. The problem therefore reduces in the bright--cavity subspace to a two-level non-Hermitian Hamiltonian. At zero detuning, $\omega_C=\omega_M$, its complex polariton eigenvalues are
%
\begin{equation}
	\widetilde E_{\rm UP/LP}
	=
	\omega_M
	-i\frac{\gamma_C+\gamma_M}{2}
	\pm
	\frac{1}{2}
	\sqrt{
		\left(\Omega_R^0\right)^2
		-
		\left(\gamma_C-\gamma_M\right)^2
	}.
	\label{eq:NHpoles}
\end{equation}
Thus, for $|\gamma_C-\gamma_M|<\Omega_R^0$, the real parts of the two homogeneous polariton eigenvalues are separated by $\Omega_{\rm NH}^{0} = \sqrt{\left(\Omega_R^0\right)^2	- \left(\gamma_C-\gamma_M\right)^2}$, while both branches have the same HWHM linewidth $(\gamma_C+\gamma_M)/2$. When $|\gamma_C-\gamma_M|>\Omega_R^0$, the square root becomes imaginary: the real parts of the two eigenvalues coalesce while their linewidths separate. The boundary $|\gamma_C-\gamma_M|=\Omega_R^0$ is the \emph{exceptional point} of the homogeneous two-mode problem, where the two complex eigenvalues and their eigenvectors coalesce\cite{Miri2019Science,Lambert2025CoherentControl,Meng2024Review}. The present work does not address exceptional point physics; all calculations below remain well within the real-energy splitting regime, with $|\gamma_C-\gamma_M|\leq0.45\,\Omega_R^0$.

Further insight into the role of the linewidths follows directly from Eq.~\ref{eq:Heff}, which can be rewritten as
\begin{equation}
	\hat H_{\rm eff}
	=
	\hat H
	-i\gamma_M\hat I
	-i(\gamma_C-\gamma_M)\ket{L}\bra{L}.
\end{equation}
This form makes clear that the eigenvectors depend only on the cavity--molecule linewidth imbalance $\gamma_C-\gamma_M$. The common term $-i\gamma_M\hat I$ gives only a uniform imaginary shift of all eigenvalues and therefore does not modify the eigenvectors. Moreover, reversing the sign of $\gamma_C-\gamma_M$ complex conjugates the Hamiltonian relative to this common shift and leaves the localization measures constructed from the magnitudes of the molecular amplitudes unchanged. The molecular localization therefore depends on $|\gamma_C-\gamma_M|$, irrespective of which bare resonance has the larger linewidth. We consequently use the linewidth-imbalance magnitude throughout the eigenvector analysis below. 

\subsubsection{Spectral and molecular-extent responses to finite linewidths}

We first examine how finite cavity and molecular linewidths modify the spectral response and the molecular localization behavior established above in the Hermitian system. In the non-Hermitian calculation, each hybrid eigenvalue is complex, so that the discrete eigenstate contributions of Eq.~\ref{eq:3} acquire finite linewidths. For visualization of the cavity-visible spectrum, we represent each hybrid mode by a Lorentzian centered at $E_n$, with HWHM $\Gamma_n$, and weighted by its photonic character $p_n$\cite{Barnes_2020,Diniz2011,Jianshu2022PRB}. The disorder-averaged spectral visibility is therefore written as
\begin{equation}
	I(\omega)
	=
	\left\langle
	\sum_n
	p_n
	\frac{\Gamma_n/\pi}
	{(\omega-E_n)^2+\Gamma_n^2}
	\right\rangle_{\mathrm{dis}}.
	\label{eq:NHspectrum}
\end{equation}
This expression is the dissipative analogue of Eq.~\ref{eq:3}, with the delta function contributions replaced by Lorentzian resonances whose widths are set by the imaginary parts of the complex eigenvalues. Equation~\ref{eq:NHspectrum} is used here as a positive modal visualization of the resonance energies, linewidths, and cavity visibility of the complex eigenmodes. Because non-Hermitian eigenmodes are generally nonorthogonal, this Lorentzian construction should not be identified with an exact eigenmode decomposition of the cavity Green's-function spectral function\cite{GardinerCollett1985,Daniel2023Plexciton}. The latter is obtained directly from
$G_{LL}(\omega)=\langle L|[\omega-\hat H_{\mathrm{eff}}]^{-1}|L\rangle$ as $-\pi^{-1}\mathrm{Im}\,G_{LL}(\omega)$. As in Sec.~II.C, the
cavity-projected response should in turn be distinguished from a measured transmission or absorption spectrum, which includes the external optical coupling geometry.

\begin{figure}[ht!]
	\centering
	\includegraphics[width=\textwidth]{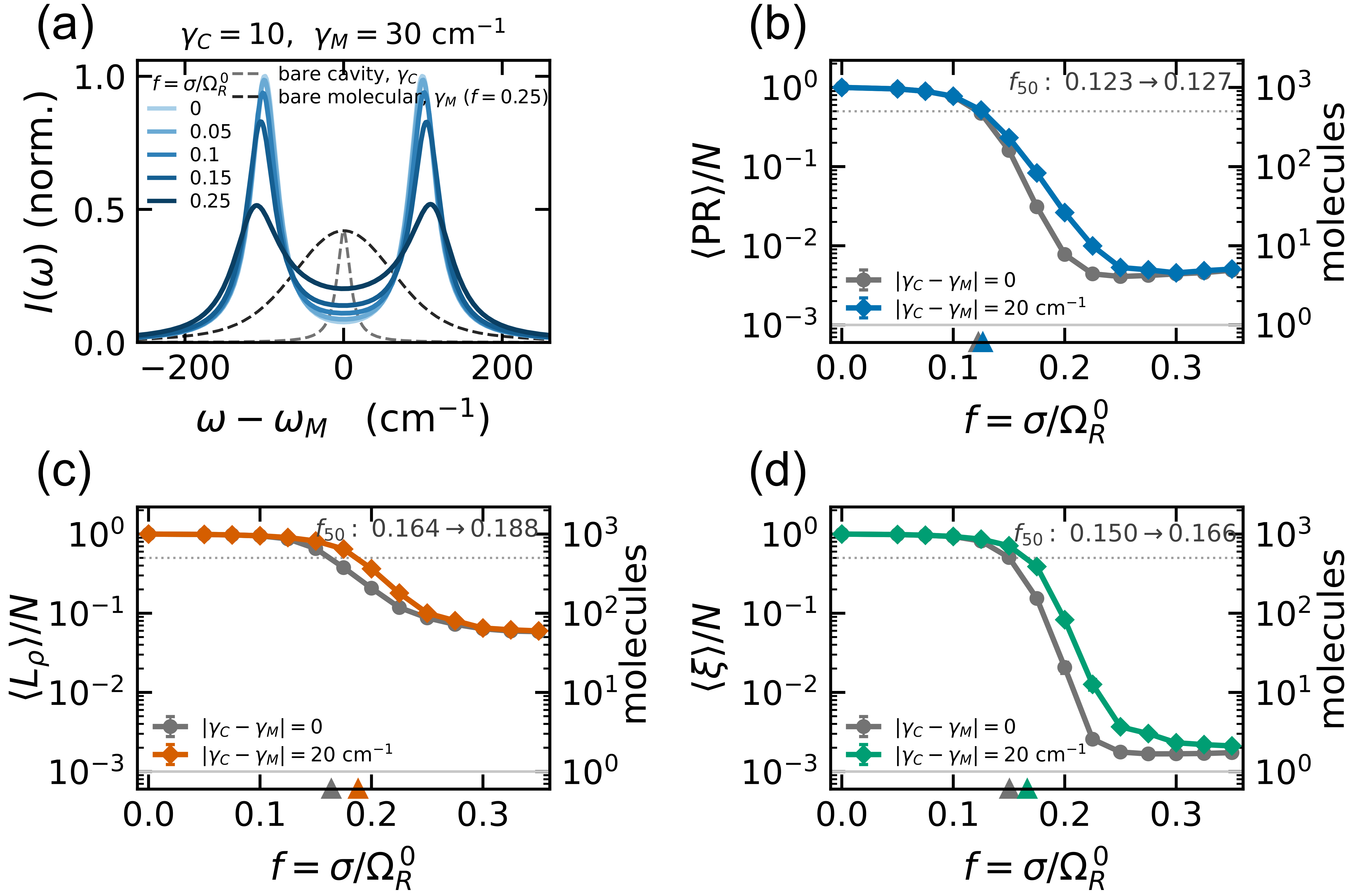}
	\caption{\footnotesize Spectral and molecular-extent response of the non-Hermitian cavity--molecule system. (a) Disorder-averaged spectral visibility for increasing static energetic disorder $f=\sigma/\Omega_R^{0}$ at fixed linewidths $\gamma_C=10~{\rm cm}^{-1}$ and $\gamma_M=30~{\rm cm}^{-1}$. The spectra are normalized to the maximum of the $f=0$ spectrum, and the same normalization factor is applied at all disorder strengths. For reference, the gray and black dashed curves denote the uncoupled cavity and molecular responses, respectively. The uncoupled molecular response is a Voigt profile combining the homogeneous molecular linewidth $\gamma_M$ with the inhomogeneous broadening at the largest disorder shown. Both reference spectra are normalized to unit maximum and scaled by a common factor of $0.45$ for visual comparison. (b)--(d) Disorder dependence of the normalized PR, $L_\rho$, and $\xi$, respectively, for the equal linewidth reference $|\gamma_C-\gamma_M|=0$ and for $|\gamma_C-\gamma_M|=20~\mathrm{cm}^{-1}$. The same eigenstate selection and averaging protocol as in Figs.~\ref{fig:fig2} and \ref{fig:fig3} is used throughout. The corresponding half-value crossover positions $f_{50}$ are marked in each panel. Error bars are standard errors of the mean over $100$ disorder realizations and are smaller than the symbols where not visible.}
	\label{fig:fig4}
\end{figure}

Figure~\ref{fig:fig4} shows the cavity-visible spectral response and the molecular localization measures for a representative non-Hermitian case with
$\gamma_C=10~{\rm cm}^{-1}$ and $\gamma_M=30~{\rm cm}^{-1}$. We choose $\gamma_M>\gamma_C$ as a representative linewidth hierarchy for molecular Fabry--P\'erot microcavities, where the molecular absorption can be broader than the cavity resonance\cite{Coles2017,Qureshi2024}. The opposite hierarchy is also encountered in other strong-coupling platforms; in plasmonic cavities, for example, Ohmic damping and radiative leakage can make the photonic resonance substantially broader than the molecular transition\cite{Torma_2015,Barnes_2020,Daniel2021Chem,Rider2022}. As shown above, however, the eigenvector-based localization measures depend only on the magnitude of the linewidth imbalance, $|\gamma_C-\gamma_M|$. The conclusions regarding molecular delocalization therefore apply equally to the opposite linewidth hierarchy, whereas the detailed spectral response in Fig.~\ref{fig:fig4}(a) corresponds to the particular choice $\gamma_M>\gamma_C$.

The spectral visibility in Fig.~\ref{fig:fig4}(a) shows the combined effects of homogeneous linewidth and static energetic disorder. At $f=0$, the LP and UP resonances remain well resolved but acquire a finite HWHM $(\gamma_C+\gamma_M)/2=20~{\rm cm}^{-1}$ due to the cavity and molecular losses. As $f$ increases, the spectral weight is progressively redistributed toward the molecular resonance near $\omega_M$, as in the Hermitian case, while the entire response is additionally broadened by the finite homogeneous linewidths. Consequently, the polariton resonances become less spectrally isolated from the disorder-broadened molecular manifold as the disorder increases, reducing the resolvability of the LP--UP doublet.
\FloatBarrier


The molecular eigenvectors, however, show a different response to the linewidth imbalance. Figures~\ref{fig:fig4}(b)--(d) compare the PR, $L_\rho$, and spectral-index correlation width, evaluated at the individual eigenstate level using the same averaging and state-selection protocol as in the Hermitian analysis, with the equal-linewidth reference, whose eigenvectors coincide with those of the Hermitian Hamiltonian. At $\sigma=0$, all three normalized measures remain equal to unity despite the unequal linewidths. This follows because finite $\gamma_C$ and $\gamma_M$ modify the complex polariton energies but do not break the permutation symmetry of the homogeneous molecular ensemble, so the molecular component remains the symmetric bright state. With increasing disorder, all three measures still undergo the pronounced decrease found in the Hermitian system, showing that the disorder-induced loss of molecular extent remains present. Its characteristic disorder scale, however, is shifted toward \emph{larger} $f$ by the linewidth imbalance. For $|\gamma_C-\gamma_M|=20~{\rm cm}^{-1}=0.1\,\Omega_R^0$, the half-value crossover shifts from $f_{50}=0.123$ to 0.127 for the PR, from 0.164 to 0.188 for $L_\rho$, and from 0.150 to 0.166 for $\xi$. The shift is modest for the PR at this linewidth imbalance but is larger for $L_\rho$ and $\xi$; all three measures move consistently toward larger static disorder. 

Thus, finite linewidth imbalance does not introduce an additional eigenstate-localization tendency in this model. Instead, the disorder-induced reduction of molecular extent occurs at larger static disorder. This contrasts with the simultaneous spectral broadening in Fig.~\ref{fig:fig4}(a): the polariton resonances become less spectrally isolated even while the eigenstate-localization crossover shifts to higher disorder. The origin of this contrasting response follows from the complex-energy structure of the non-Hermitian eigenvectors, discussed in Sec.~II.E.2.

\subsubsection{Origin of the delayed localization crossover}

The shift of the localization crossover toward larger disorder with increasing linewidth imbalance can be understood directly from the molecular amplitudes of the non-Hermitian eigenvectors. Projecting the eigenvalue equation for Eq.~\ref{eq:Heff} onto the $j$th molecular state gives the non-Hermitian generalization of Eq.~\ref{eq:8},
\begin{equation}
	a_j^{(n)}
	=
	\frac{g_c\ell_n}
	{\widetilde E_n-\widetilde\omega_j}
	=
	\frac{g_c\ell_n}
	{(E_n-\omega_j)-i(\Gamma_n-\gamma_M)} .
	\label{eq:NHamp}
\end{equation}
The molecular weight is therefore controlled not only by the real-energy separation $E_n-\omega_j$, as in the Hermitian problem, but also by the difference between the linewidth of the hybrid state and the molecular linewidth.

For the present effective Hamiltonian, this imaginary energy separation can be related directly to the photonic weight of the eigenstate. For a normalized right eigenvector, taking the imaginary part of $\langle\psi_n|\hat H_{\mathrm{eff}}|\psi_n\rangle =\widetilde E_n\langle\psi_n|\psi_n\rangle$ gives
$\Gamma_n=p_n\gamma_C+(1-p_n)\gamma_M,$ and therefore $\Gamma_n-\gamma_M=p_n(\gamma_C-\gamma_M)$. Thus, predominantly molecular states with $p_n\simeq0$ retain linewidths close to $\gamma_M$, whereas an optically visible polariton acquires a finite separation from the molecular manifold along the imaginary energy direction. In the homogeneous resonant limit, where $p_n=1/2$, this separation has magnitude $|\gamma_C-\gamma_M|/2$. Substituting this into Eq.~\ref{eq:NHamp} gives
\begin{equation}
	\left|a_j^{(n)}\right|^2
	=
	\frac{
		g_c^2|\ell_n|^2
	}{
		(E_n-\omega_j)^2
		+
		p_n^2(\gamma_C-\gamma_M)^2
	} .
	\label{eq:NHamp2}
\end{equation}

When $\gamma_C=\gamma_M$, the imaginary-energy term in Eq.~\ref{eq:NHamp2} vanishes and the molecular weighting reduces to the Hermitian form discussed above. A molecular transition lying very close to the real part of the polariton energy can then acquire a disproportionately large amplitude, allowing static disorder to concentrate the molecular component onto only a few near-resonant molecules. For finite linewidth imbalance, the additional term $p_n^2(\gamma_C-\gamma_M)^2$ prevents the denominator from becoming arbitrarily small for an optically visible state. The disproportionate enhancement of an individual near-resonant transition is therefore suppressed, allowing the molecular weight to remain distributed over a broader set of transitions and shifting the loss of molecular extent to larger disorder. Even when an accidental molecular transition is exactly resonant with the real part of the polariton energy, $E_n=\omega_j$, the complex energy denominator remains finite whenever $p_n>0$ and $\gamma_C\neq\gamma_M$.

Further, Eq.\ref{eq:NHamp2} has a Lorentzian form with HWHM
$p_n|\gamma_C-\gamma_M|$ when viewed as a function of the  real-energy separation $E_n-\omega_j$. As $|\gamma_C-\gamma_M|$ increases, this broadens the molecular amplitude profile in energy and suppresses the strong weighting of individual near-resonant molecules that promotes localization in the Hermitian problem. This provides a direct explanation for the shift observed in Fig.~\ref{fig:fig4}, suggesting that the linewidth imbalance delays the disorder-induced reduction of molecular extent. \si{The corresponding increase of the minimum complex-energy denominator and the reduction of the largest single-molecule population are shown directly in SI Sec.~S11 and Fig.~S13.}


\subsubsection{Localization crossover in the disorder--linewidth plane}

\begin{figure}[ht!]
	\centering
	\includegraphics[width=\textwidth]{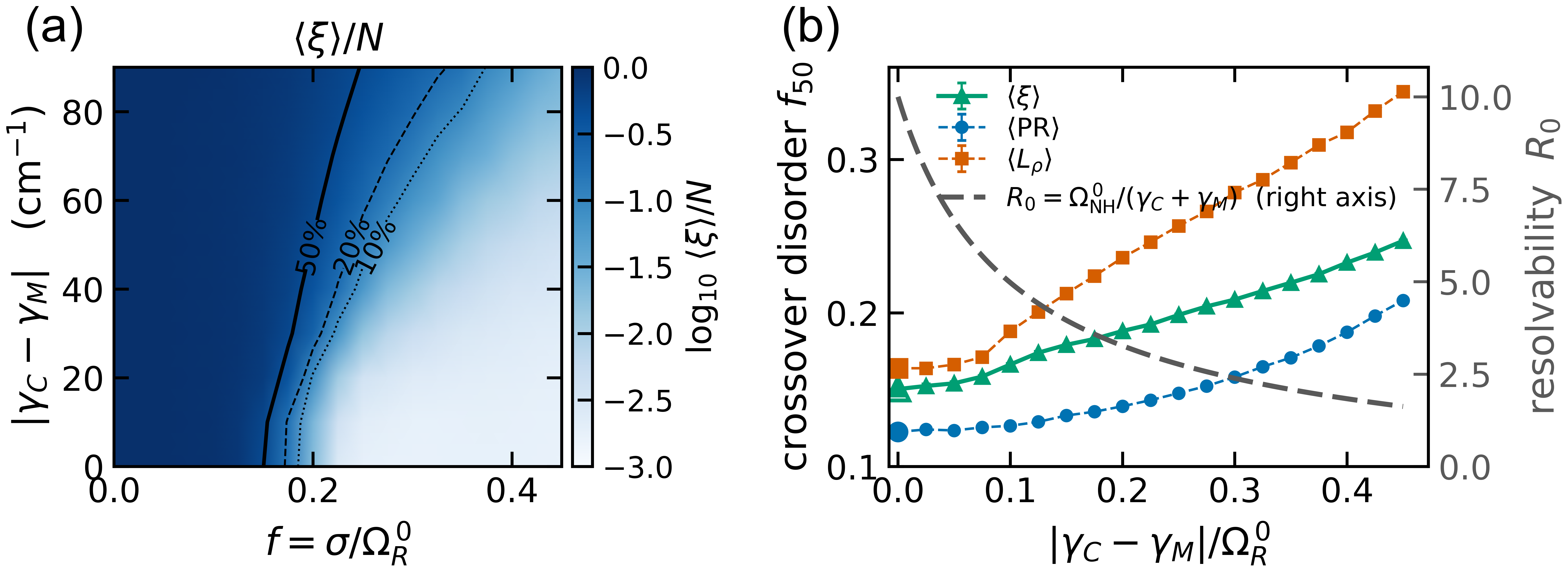}
	\caption{\footnotesize Disorder--linewidth dependence of the molecular localization crossover and spectral resolvability. (a) Color map of the normalized disorder-averaged spectral-index correlation width, $\langle\xi\rangle/N$, as a function of the static energetic disorder $f=\sigma/\Omega_R^{0}$ and linewidth imbalance $|\gamma_C-\gamma_M|$. Here $\gamma_C=10~{\rm cm}^{-1}$ is held fixed while $\gamma_M$ is varied. The color scale is logarithmic, and the contours mark the normalized correlation-width levels 0.5, 0.2, and 0.1. (b) (Left axis): half-value crossover $f_{50}$ for the PR, $L_\rho$, and $\xi$ as a function of $|\gamma_C-\gamma_M|/\Omega_R^0$. Error bars are standard deviations obtained from 300 bootstrap resamples of the disorder realizations and are smaller than the symbols where not visible. (Right axis): The homogeneous ($\sigma=0$) spectral resolvability $R_0=\Omega_{\mathrm{NH}}^0/(\gamma_C+\gamma_M)$ for the same linewidth sweep. All calculations use $N=1000$, $\Omega_R^0=200~\mathrm{cm}^{-1}$, 100 disorder realizations, and the visibility criterion $p_n\geq0.1p_{\max}$. Dependence on the crossover level and visibility threshold is examined in \si{SI Sec.~S10 and Fig.~S12.}}
\label{fig:fig5}
\end{figure}

Having identified the microscopic origin of the delayed localization crossover, we now examine how this effect evolves over a broader range of linewidth imbalance. Figure~\ref{fig:fig5}(a) maps the normalized spectral-index correlation width, $\langle\xi\rangle/N$, as a function of the static disorder $f=\sigma/\Omega_R^0$ and the linewidth imbalance $|\gamma_C-\gamma_M|$. The contours mark the normalized correlation-width levels $\langle\xi\rangle/N=0.5$, 0.2, and 0.1. The contours show an overall shift toward larger $f$ as $|\gamma_C-\gamma_M|$ increases, indicating that progressively stronger energetic disorder is required to produce the same reduction of molecular extent; in other words, the cavity-visible eigenstates become more tolerant to static disorder.

The same trend is obtained from the other localization measures. Figure~\ref{fig:fig5}(b) compares the half-value crossover $f_{50}$ extracted from the PR, $L_\rho$, and $\xi$ over $0\leq|\gamma_C-\gamma_M|\leq90~{\rm cm}^{-1} =0.45\,\Omega_R^0$. Across this range, $f_{50}$ increases from approximately $0.123$ to $0.208$ for the PR ($\sim 70\%$), from $0.164$ to $0.344$ for $L_\rho$ ($\sim 110\%$), and from $0.150$ to $0.247$ for $\xi$ ($\sim 64\%$). The magnitude of the shift differs among the measures, as expected from their different weighting of the molecular eigenvectors, but its direction is the same in every case: increasing linewidth imbalance increases the amount of static disorder that can be tolerated before a given fraction of the homogeneous molecular extent is lost.

This behavior is not specific to the half-value localization crossover $f_{50}$. The corresponding $f_{20}$ and $f_{10}$ values, defined by normalized molecular extent levels of 0.2 and 0.1, respectively, show the same overall shift toward larger disorder as $|\gamma_C-\gamma_M|$ increases, and the trend persists over a broad range of photonic-visibility thresholds (\si{SI Sec.~S10 and Fig.~S12}). The quantitative crossover position retains some dependence on the molecular-extent measure, fractional crossover level, and state-selection threshold; we therefore report the calculated crossover values directly rather than assigning a universal empirical functional form.

\FloatBarrier

The increased disorder tolerance of the molecular eigenvectors does not imply improved spectroscopic strong coupling. To make this distinction explicit, Fig.~\ref{fig:fig5}(b) also shows the homogeneous ($\sigma=0$) spectral resolvability associated with the same linewidth parameters. Using the real-part separation of the homogeneous non-Hermitian polariton eigenvalues,
$\Omega_{\rm NH}^{0} =[(\Omega_R^0)^2-(\gamma_C-\gamma_M)^2]^{1/2}$, and noting that both homogeneous polariton branches have HWHM $(\gamma_C+\gamma_M)/2$, their common FWHM is $\gamma_C+\gamma_M$. We therefore define the corresponding
spectral resolvability as
\begin{equation}
	\mathcal{R}_0
	=
	\frac{\Omega_{\rm NH}^{0}}
	{\gamma_C+\gamma_M}.
	\label{eq:resolvability}
\end{equation}
Over the linewidth range considered, $\mathcal{R}_0$ decreases from $10$ to approximately $1.62$, a reduction of approximately 84\%, while the real-part polariton separation $\Omega_{\mathrm{NH}}^0$ decreases only from 200 to $178.6~\mathrm{cm}^{-1}$, or by approximately 11\%. The pronounced loss of spectral resolvability therefore arises predominantly from the increasing linewidths rather than from a collapse of the polariton energy separation. At the same time, the half-value localization crossover shifts toward larger static disorder by approximately $64$--$110\%$, depending on the localization measure. Thus, increasing linewidth imbalance can substantially degrade spectral resolvability while allowing the molecular component of the cavity-visible eigenstates to retain its extent up to considerably stronger energetic disorder.

Note that this opposing response should not be interpreted as implying that dissipation is always favorable for polaritonic phenomena. Larger linewidths shorten the lifetime of the hybrid excitations and reduce their spectral resolvability, while the present eigenstate analysis does not address dynamical processes that depend simultaneously on lifetime, transport, relaxation, and coherence. Rather, the result shows that spectral resolvability and eigenstate-level molecular delocalization probe distinct properties of an open cavity--molecule system: linewidth imbalance can delay the disorder-induced loss of molecular extent even while the polariton spectrum becomes progressively less well resolved.\\

Together with the Hermitian results, this highlights a broader distinction between spectroscopic signatures of strong coupling and the molecular extent of the corresponding cavity-visible eigenstates. Static energetic disorder can strongly reduce molecular delocalization while a resolved LP--UP doublet remains visible and its apparent splitting can even increase, whereas linewidth imbalance reduces spectral resolvability but shifts the localization crossover toward larger disorder. Thus, the spectral appearance of the polariton doublet alone does not determine the molecular extent of the underlying eigenstates.

\subsection{Experimental Implications}
The distinction established above between spectral visibility and eigenstate-level molecular delocalization is important when relating the present results to experiment. The crossover identified here characterizes how broadly the molecular component of an optically visible polariton eigenstate is distributed, and need not coincide with changes in other observables. Whether this loss of molecular extent is relevant to a particular experiment therefore depends on whether the measured process is sensitive primarily to an individual polariton eigenstate, to the photon-weighted bright sector, or to transport and dynamics involving many states. Our bright-sector analysis illustrates this distinction directly: deep in the localized regime, the photon-weighted bright-sector PR can increase because different localized eigenstates populate different molecules, even while $L_\rho$ and the correlation width show no corresponding recovery of molecular extent. Thus, a large number of molecules sampled collectively by the cavity-accessible sector does not necessarily imply that the molecular component of any individual polariton eigenstate is similarly extended. This distinction is particularly relevant when connecting spectral measurements to experiments in which polariton delocalization has been implicated in nonlinear optical response, ultrafast energy redistribution, and coherent propagation\cite{Xiang2019,Russo2024,Pandya2021}. Such observables need not follow the same disorder dependence as either the linear polariton spectrum or a single-eigenstate localization measure.

The non-Hermitian results add a complementary consideration when comparing experiments with different cavity losses or molecular homogeneous linewidths. A broader polariton spectrum, or a cavity with a lower quality factor, does not by itself imply a more localized molecular component. In the present model, a linewidth contribution common to the cavity and molecular excitations changes the resonance lifetimes and spectral resolution without modifying the eigenvectors, whereas the cavity--molecule linewidth imbalance changes the disorder dependence of the molecular extent. Consequently, spectral linewidths or cavity quality factors should not be used alone as proxies for molecular delocalization when comparing cavities, materials, or strong-coupling platforms with different loss channels. This distinction is particularly relevant across systems in which the dominant broadening mechanism can change, for example between molecular Fabry--P\'erot cavities and more strongly damped plasmonic structures\cite{Torma_2015,Barnes_2020,Daniel2021Chem,Rider2022}.

Quantitative comparison with experiment should therefore specify the observable and disorder model used to define the crossover and, when dissipation is important, the corresponding cavity and molecular linewidths. For the Gaussian ensemble considered here, retaining approximately one half of the homogeneous molecular extent requires $\Omega_R^0/\sigma$ of order $7$, consistent with participation-based analyses showing that substantial collective coupling can be required to overcome static energetic disorder.\cite{Wei2025ChemSci} The numerical coefficient is nevertheless not universal: bounded disorder distributions shift the crossover substantially, while different localization measures and retention levels probe different stages of the same continuous crossover. The present analysis applies most directly to electronic strong coupling in the weak-excitation regime. The range of collective splittings and optical linewidths considered here is representative of organic molecular microcavities, where energy scales of tens to hundreds of wavenumbers are commonly encountered.\cite{Coles2017,Qureshi2024} The same eigenvector construction can formally be applied to the linear infrared response under vibrational strong coupling, but extending the present criterion to vibrational-polariton chemistry or thermal dynamics requires additional physics absent from the single-excitation Tavis--Cummings description, including anharmonicity, relaxation, coupling to dynamical baths, and temperature-dependent molecular fluctuations.\cite{HerreraSpano2016,HerreraSpano2018} Accordingly, the transferable point is not a single numerical crossover value, but the need to distinguish the spectroscopic visibility of a polariton from the molecular extent of the states involved in the particular process being measured.

\section{Conclusions}
In conclusion, our results show that a well-resolved polariton splitting is not, by itself, sufficient to establish that the molecular component of a polariton remains delocalized over the ensemble. Using a disordered Tavis--Cummings model, we show that static energetic disorder can strongly reduce the molecular extent of the optically visible polariton states even while a clearly resolved LP--UP doublet remains visible and its apparent splitting can increase. By quantifying the molecular extent of the optically visible states using the participation ratio, the molecular coherence measure $L_\rho$, and the spectral-index correlation width, we identify a continuous crossover in eigenstate-level molecular delocalization, from an extended bright-state-like molecular component to one concentrated over a much smaller fraction of the ensemble. Finite-size scaling shows that the crossover position approaches a stable large-$N$ scale over the accessible ensemble sizes; we therefore characterize the observed behavior as a finite-size crossover rather
than infer a sharp thermodynamic transition. For Gaussian disorder, retaining approximately one half of the homogeneous molecular extent requires a bare collective coupling $\Omega_R^0/\sigma$ of order $7$, although the precise numerical scale depends on the localization measure and, importantly, on the form of the disorder distribution.

Including finite cavity and molecular linewidths through an effective non-Hermitian Hamiltonian reveals a contrasting role of linewidth imbalance. A common linewidth contribution changes the complex energies without modifying the molecular eigenvectors, whereas a finite cavity--molecule linewidth imbalance suppresses the disproportionate weighting of near-resonant molecular transitions and shifts the disorder-induced crossover toward larger static disorder. Over the range studied, this increased tolerance of the molecular extent occurs even as the polariton spectrum becomes substantially less well resolved. The implication is therefore not that dissipation is favorable, but that spectroscopic signatures of strong coupling and eigenstate-level molecular delocalization probe distinct properties of an open molecular polariton system and need not evolve in parallel. This distinction is especially relevant to organic and molecular microcavities, where collective coupling coexists with substantial energetic disorder, homogeneous broadening, and vibronic structure\cite{Houdre1996,Cwik2016,HerreraSpano2018,Arkajit2023Rev}. The present work provides a transparent electronic baseline for separating these effects and motivates future extensions incorporating vibronic coupling, intermolecular interactions, dynamic disorder, and multimode cavity structure\cite{HerreraSpano2016,HerreraSpano2018,Ribeiro2018}.

\section*{Supplementary Material}
See the Supporting Information (SI) for spectral construction and numerical checks (Sec.~S1; Figs.~S1--S2); calibration of the molecular extent measures, molecular amplitude profile, and analysis protocols (Secs.~S2--S5; Figs.~S3--S6); finite-size, disorder distribution, fixed-window, and energy-space correlation analyses (Secs.~S6--S9; Figs.~S7--S11); robustness and amplitude-level interpretation of the non-Hermitian crossover (Secs.~S10--S11; Figs.~S12--S13).

\section*{Acknowledgements}
A.S. acknowledges research fellowship from STINT, the Swedish Foundation for International Cooperation in Research and Higher Education (PS2024-9486). T.P. acknowledges research funding from the Swedish Foundation of Strategic Research (IS24-0005) and the Olle Engkvists Foundation (235-0422).
\section*{Author Declarations}
\subsection*{Conflict of Interest}
The authors have no conflicts to disclose.
\section*{Data Availability}
The data that support the findings of this article are available from the corresponding author upon reasonable request.

\bibliography{refs}%

@article{Tavis1968,
  title = {Exact Solution for an $N$-Molecule---Radiation-Field Hamiltonian},
  author = {Tavis, Michael and Cummings, Frederick W.},
  journal = {Phys. Rev.},
  volume = {170},
  issue = {2},
  pages = {379--384},
  numpages = {0},
  year = {1968},
  month = {Jun},
  publisher = {American Physical Society},
  doi = {10.1103/PhysRev.170.379},
  url = {https://link.aps.org/doi/10.1103/PhysRev.170.379}
}

@article{Dicke1954,
  title        = {Coherence in Spontaneous Radiation Processes},
  author       = {Dicke, R. H.},
  journal      = {Phys. Rev.},
  volume       = {93},
  pages        = {99--110},
  year         = {1954},
  doi          = {10.1103/PhysRev.93.99}
}

@article{GardinerCollett1985,
  title        = {Input and output in damped quantum systems: Quantum stochastic differential equations and the master equation},
  author       = {Gardiner, C. W. and Collett, M. J.},
  journal      = {Phys. Rev. A},
  volume       = {31},
  pages        = {3761--3774},
  year         = {1985},
  doi          = {10.1103/PhysRevA.31.3761}
}

@book{Carmichael1993,
  title        = {An Open Systems Approach to Quantum Optics},
  author       = {Carmichael, Howard},
  publisher    = {Springer},
  year         = {1993},
  doi          = {10.1007/978-3-540-47620-7}
}

@article{Anderson1958,
  title        = {Absence of Diffusion in Certain Random Lattices},
  author       = {Anderson, P. W.},
  journal      = {Phys. Rev.},
  volume       = {109},
  pages        = {1492--1505},
  year         = {1958},
  doi          = {10.1103/PhysRev.109.1492}
}

@article{Diniz2011,
  title = {Strongly coupling a cavity to inhomogeneous ensembles of emitters: Potential for long-lived solid-state quantum memories},
  author = {Diniz, I. and Portolan, S. and Ferreira, R. and G\'erard, J. M. and Bertet, P. and Auff\`eves, A.},
  journal = {Phys. Rev. A},
  volume = {84},
  issue = {6},
  pages = {063810},
  numpages = {9},
  year = {2011},
  month = {Dec},
  publisher = {American Physical Society},
  doi = {10.1103/PhysRevA.84.063810},
  url = {https://link.aps.org/doi/10.1103/PhysRevA.84.063810}
}

@article{Michetti2005,
  title = {Polariton states in disordered organic microcavities},
  author = {Michetti, P. and La Rocca, G. C.},
  journal = {Phys. Rev. B},
  volume = {71},
  issue = {11},
  pages = {115320},
  numpages = {7},
  year = {2005},
  month = {Mar},
  publisher = {American Physical Society},
  doi = {10.1103/PhysRevB.71.115320},
  url = {https://link.aps.org/doi/10.1103/PhysRevB.71.115320}
}

@article{Whittaker1998,
  title = {What Determines Inhomogeneous Linewidths in Semiconductor Microcavities?},
  author = {Whittaker, D. M.},
  journal = {Phys. Rev. Lett.},
  volume = {80},
  issue = {21},
  pages = {4791--4794},
  numpages = {0},
  year = {1998},
  month = {May},
  publisher = {American Physical Society},
  doi = {10.1103/PhysRevLett.80.4791},
  url = {https://link.aps.org/doi/10.1103/PhysRevLett.80.4791}
}

@article{Sebastian2022,
  author  = {Gera, Tarun and Sebastian, K. L.},
  title   = {Effects of disorder on polaritonic and dark states in a cavity using the disordered Tavis--Cummings model},
  journal = {The Journal of Chemical Physics},
  volume  = {156},
  pages   = {194304},
  year    = {2022},
  doi     = {10.1063/5.0086027},
  eprint  = {https://pubs.aip.org/aip/jcp/article-pdf/doi/10.1063/5.0086027/20855832/194304_1_5.0086027.pdf}
}

@article{Grochol2008,
  title = {Microcavity polaritons in disordered exciton lattices},
  author = {Grochol, Michal and Piermarocchi, Carlo},
  journal = {Phys. Rev. B},
  volume = {78},
  issue = {3},
  pages = {035323},
  numpages = {7},
  year = {2008},
  month = {Jul},
  publisher = {American Physical Society},
  doi = {10.1103/PhysRevB.78.035323},
  url = {https://link.aps.org/doi/10.1103/PhysRevB.78.035323}
}

@article{Murphy2011,
  author  = {Murphy, N. C. and Wortis, R. and Atkinson, W. A.},
  title   = {Generalized inverse participation ratio as a possible measure of localization for interacting systems},
  journal = {Physical Review B},
  volume  = {83},
  pages   = {184206},
  year    = {2011},
  doi     = {10.1103/PhysRevB.83.184206}
}

@article{KramerMacKinnon1993,
  author  = {Kramer, Bernhard and MacKinnon, Alan},
  title   = {Localization: theory and experiment},
  journal = {Reports on Progress in Physics},
  volume  = {56},
  number  = {12},
  pages   = {1469--1564},
  year    = {1993},
  doi     = {10.1088/0034-4885/56/12/001}
}

@article{Bondarenko2020,
   author = {Bondarenko, Anna S. and Jansen, Thomas L. C. and Knoester, Jasper},
   title = {Exciton localization in tubular molecular aggregates: Size effects and optical response},
   journal = {The Journal of Chemical Physics},
   volume = {152},
   number = {19},
   pages = {194302},
   year = {2020},
   month = {05},
   issn = {0021-9606},
   doi = {10.1063/5.0008688},
   url = {https://doi.org/10.1063/5.0008688},
   eprint = {https://pubs.aip.org/aip/jcp/article-pdf/doi/10.1063/5.0008688/15575519/194302_1_online.pdf}
}

@article{Jiang2023,
author = {Jiang, Tong and Ren, Jiajun and Shuai, Zhigang},
title = {Unified Definition of Exciton Coherence Length for Exciton–Phonon Coupled Molecular Aggregates},
journal = {The Journal of Physical Chemistry Letters},
volume = {14},
number = {19},
pages = {4541-4547},
year = {2023},
doi = {10.1021/acs.jpclett.3c00812},
    note ={PMID: 37159446},

URL = { 
    
        https://doi.org/10.1021/acs.jpclett.3c00812
    
    

},
eprint = { 
    
        https://doi.org/10.1021/acs.jpclett.3c00812
    
    

}

}

@article{Meier1997,
    author = {Meier, T. and Zhao, Y. and Chernyak, V. and Mukamel, S.},
    title = {Polarons, localization, and excitonic coherence in superradiance of biological antenna complexes},
    journal = {The Journal of Chemical Physics},
    volume = {107},
    number = {10},
    pages = {3876-3893},
    year = {1997},
    month = {09},
    issn = {0021-9606},
    doi = {10.1063/1.474746},
    url = {https://doi.org/10.1063/1.474746},
    eprint = {https://pubs.aip.org/aip/jcp/article-pdf/107/10/3876/19035832/3876_1_online.pdf},
}

@article{Scholes2006,
  author  = {Scholes, Gregory D. and Rumbles, Garry},
  title   = {Excitons in Nanoscale Systems},
  journal = {Nature Materials},
  volume  = {5},
  pages   = {683--696},
  year    = {2006},
  doi     = {10.1038/nmat1710}
}

@article{Kuhn1997,
   author = {Kühn, Oliver and Sundström, Villy},
   title = {Pump–probe spectroscopy of dissipative energy transfer dynamics in photosynthetic antenna complexes: A density matrix approach},
   journal = {The Journal of Chemical Physics},
   volume = {107},
   number = {11},
   pages = {4154-4164},
   year = {1997},
   month = {09},
   issn = {0021-9606},
   doi = {10.1063/1.474803},
   url = {https://doi.org/10.1063/1.474803},
   eprint = {https://pubs.aip.org/aip/jcp/article-pdf/107/11/4154/19289278/4154_1_online.pdf}
}

@article{Pullerits2001,
  author  = {Dahlbom, M. and Pullerits, T. and Mukamel, S. and Sundstr{\"o}m, V.},
  title   = {Exciton Delocalization in the {B850} Light-Harvesting Complex: Comparison of Different Measures},
  journal = {The Journal of Physical Chemistry B},
  volume  = {105},
  number  = {23},
  pages   = {5515--5524},
  year    = {2001},
  doi     = {10.1021/jp004496i}
}

@article{Daniel2023Plexciton,
    author = {Finkelstein-Shapiro, Daniel and Mante, Pierre-Adrien and Balci, Sinan and Zigmantas, Donatas and Pullerits, Tõnu},
    title = {Non-Hermitian Hamiltonians for linear and nonlinear optical response: A model for plexcitons},
    journal = {The Journal of Chemical Physics},
    volume = {158},
    number = {10},
    pages = {104104},
    year = {2023},
    month = {03},
    issn = {0021-9606},
    doi = {10.1063/5.0130287},
    url = {https://doi.org/10.1063/5.0130287},
    eprint = {https://pubs.aip.org/aip/jcp/article-pdf/doi/10.1063/5.0130287/18060945/104104_1_5.0130287.pdf}
}

@article{Jonathan2022,
    author = {McTague, Jonathan and Foley, Jonathan J., IV},
    title = {Non-Hermitian cavity quantum electrodynamics–configuration interaction singles approach for polaritonic structure with ab initio molecular Hamiltonians},
    journal = {The Journal of Chemical Physics},
    volume = {156},
    number = {15},
    pages = {154103},
    year = {2022},
    month = {04},
    issn = {0021-9606},
    doi = {10.1063/5.0091953},
    url = {https://doi.org/10.1063/5.0091953},
    eprint = {https://pubs.aip.org/aip/jcp/article-pdf/doi/10.1063/5.0091953/16540152/154103_1_online.pdf},
}

@article{Miri2019Science,
author = {Mohammad-Ali Miri  and Andrea Alù },
title = {Exceptional points in optics and photonics},
journal = {Science},
volume = {363},
number = {6422},
pages = {eaar7709},
year = {2019},
doi = {10.1126/science.aar7709},
URL = {https://www.science.org/doi/abs/10.1126/science.aar7709},
eprint = {https://www.science.org/doi/pdf/10.1126/science.aar7709}
}

@article{Lambert2025CoherentControl,
  title={Coherent control of magnon-polaritons using an exceptional point},
  author={Lambert, N.J. and Schumer, A. and Longdell, J.J. and Rotter, S. and Schwefel, H. G. L.},
  journal={Nature Physics},
  volume={21},
  pages={1570--1577},
  year={2025},
  doi={10.1038/s41567-025-02998-3},
  url={https://doi.org/10.1038/s41567-025-02998-3}
}

@article{Meng2024Review,
    author = {Meng, Haiyu and Ang, Yee Sin and Lee, Ching Hua},
    title = {Exceptional points in non-Hermitian systems: Applications and recent developments},
    journal = {Applied Physics Letters},
    volume = {124},
    number = {6},
    pages = {060502},
    year = {2024},
    month = {02},
    issn = {0003-6951},
    doi = {10.1063/5.0183826},
    url = {https://doi.org/10.1063/5.0183826},
    eprint = {https://pubs.aip.org/aip/apl/article-pdf/doi/10.1063/5.0183826/19956853/060502_1_5.0183826.pdf}
}

@book{Moiseyev2011,
  author    = {Moiseyev, Nimrod},
  title     = {Non-Hermitian Quantum Mechanics},
  publisher = {Cambridge University Press},
  address   = {Cambridge},
  year      = {2011},
  doi       = {10.1017/CBO9780511976186},
  isbn      = {9780521889728}
}

@article{Wei2025ChemSci,
    author = {Liu, Tianlin and Yin, Guoxin and Xiong, Wei},
    title = {Unlocking delocalization: how much coupling strength is required to overcome energy disorder in molecular polaritons?},
    journal = {Chemical Science},
    volume = {16},
    number = {11},
    pages = {4676-4683},
    year = {2025},
    month = {03},
    issn = {2041-6520},
    doi = {10.1039/d4sc07053d},
    url = {https://doi.org/10.1039/d4sc07053d},
    eprint = {https://pubs.rsc.org/sc/article-pdf/16/11/4676/10050387/d4sc07053d.pdf}
}

@article{Jianshu2022PRB,
  title = {Unusual dynamical properties of disordered polaritons in microcavities},
  author = {Engelhardt, Georg and Cao, Jianshu},
  journal = {Phys. Rev. B},
  volume = {105},
  issue = {6},
  pages = {064205},
  numpages = {19},
  year = {2022},
  month = {Feb},
  publisher = {American Physical Society},
  doi = {10.1103/PhysRevB.105.064205},
  url = {https://link.aps.org/doi/10.1103/PhysRevB.105.064205}
}

@article{Barnes_2020,
doi = {10.1088/2040-8986/ab7b01},
url = {https://doi.org/10.1088/2040-8986/ab7b01},
year = {2020},
month = {jun},
publisher = {IOP Publishing},
volume = {22},
number = {7},
pages = {073501},
author = {Barnes, William L and Horsley, Simon A R and Vos, Willem L},
title = {Classical antennas, quantum emitters, and densities of optical states},
journal = {Journal of Optics}
}

@article{Gerrit2024JCP,
    author = {Sokolovskii, Ilia and Groenhof, Gerrit},
    title = {Non-Hermitian molecular dynamics simulations of exciton–polaritons in lossy cavities},
    journal = {The Journal of Chemical Physics},
    volume = {160},
    number = {9},
    pages = {092501},
    year = {2024},
    month = {03},
    issn = {0021-9606},
    doi = {10.1063/5.0188613},
    url = {https://doi.org/10.1063/5.0188613},
    eprint = {https://pubs.aip.org/aip/jcp/article-pdf/doi/10.1063/5.0188613/19702666/092501_1_5.0188613.pdf}
}

@article{HerreraSpano2018,
author = {Herrera, Felipe and Spano, Frank C.},
title = {Theory of Nanoscale Organic Cavities: The Essential Role of Vibration-Photon Dressed States},
journal = {ACS Photonics},
volume = {5},
number = {1},
pages = {65-79},
year = {2018},
doi = {10.1021/acsphotonics.7b00728},

URL = { 
    
        https://doi.org/10.1021/acsphotonics.7b00728
    
    

},
eprint = { 
    
        https://doi.org/10.1021/acsphotonics.7b00728
    
    

}

}

@article{HerreraSpano2016,
  title = {Cavity-Controlled Chemistry in Molecular Ensembles},
  author = {Herrera, Felipe and Spano, Frank C.},
  journal = {Phys. Rev. Lett.},
  volume = {116},
  issue = {23},
  pages = {238301},
  numpages = {6},
  year = {2016},
  month = {Jun},
  publisher = {American Physical Society},
  doi = {10.1103/PhysRevLett.116.238301},
  url = {https://link.aps.org/doi/10.1103/PhysRevLett.116.238301}
}

@article{Cwik2016,
  title = {Excitonic spectral features in strongly coupled organic polaritons},
  author = {\ifmmode \acute{C}\else \'{C}\fi{}wik, Justyna A. and Kirton, Peter and De Liberato, Simone and Keeling, Jonathan},
  journal = {Phys. Rev. A},
  volume = {93},
  issue = {3},
  pages = {033840},
  numpages = {12},
  year = {2016},
  month = {Mar},
  publisher = {American Physical Society},
  doi = {10.1103/PhysRevA.93.033840},
  url = {https://link.aps.org/doi/10.1103/PhysRevA.93.033840}
}

@article{Coles2017,
author = {Coles, David M. and Chen, Qiang and Flatten, Lucas C. and Smith, Jason M. and M{\"u}llen, Klaus and Narita, Akimitsu and Lidzey, David G.},
title = {Strong Exciton–Photon Coupling in a Nanographene Filled Microcavity},
journal = {Nano Letters},
volume = {17},
number = {9},
pages = {5521-5525},
year = {2017},
doi = {10.1021/acs.nanolett.7b02211},
    note ={PMID: 28829137},

URL = { 
    
        https://doi.org/10.1021/acs.nanolett.7b02211
    
    

},
eprint = { 
    
        https://doi.org/10.1021/acs.nanolett.7b02211
    
    

}

}

@article{Qureshi2024,
author = {Qureshi, Hassan A. and Papachatzakis, Michael A. and Abdelmagid, Ahmed Gaber and Salomäki, Mikko and Mäkilä, Ermei and Tuomi, Oskar and Siltanen, Olli and Daskalakis, Konstantinos S.},
title = {Giant Rabi Splitting and Polariton Photoluminescence in an all Solution-Deposited Dielectric Microcavity},
journal = {Advanced Optical Materials},
volume = {13},
number = {16},
pages = {2500155},
doi = {https://doi.org/10.1002/adom.202500155},
url = {https://advanced.onlinelibrary.wiley.com/doi/abs/10.1002/adom.202500155},
eprint = {https://advanced.onlinelibrary.wiley.com/doi/pdf/10.1002/adom.202500155},
year = {2025}
}

@article{ElGanainy2018,
  title   = {Non-Hermitian physics and PT symmetry},
  author  = {El-Ganainy, Ramy and Makris, Konstantinos G. and Khajavikhan, Mercedeh and Musslimani, Ziad H. and Rotter, Stefan and Christodoulides, Demetrios N.},
  journal = {Nature Physics},
  volume  = {14},
  pages   = {11--19},
  year    = {2018},
  doi     = {10.1038/nphys4323}
}

@article{Ebbesen2016,
author = {Ebbesen, Thomas W.},
title = {Hybrid Light–Matter States in a Molecular and Material Science Perspective},
journal = {Accounts of Chemical Research},
volume = {49},
number = {11},
pages = {2403-2412},
year = {2016},
doi = {10.1021/acs.accounts.6b00295},
    note ={PMID: 27779846},

URL = { 
    
        https://doi.org/10.1021/acs.accounts.6b00295
    
    

},
eprint = { 
    
        https://doi.org/10.1021/acs.accounts.6b00295
    
    

}

}

@article{Ribeiro2018,
    author = {Ribeiro, Raphael F. and Martínez-Martínez, Luis A. and Du, Matthew and Campos-Gonzalez-Angulo, Jorge and Yuen-Zhou, Joel},
    title = {Polariton chemistry: controlling molecular dynamics with optical cavities},
    journal = {Chemical Science},
    volume = {9},
    number = {30},
    pages = {6325-6339},
    year = {2018},
    month = {08},
    issn = {2041-6520},
    doi = {10.1039/c8sc01043a},
    url = {https://doi.org/10.1039/c8sc01043a},
    eprint = {https://pubs.rsc.org/sc/article-pdf/9/30/6325/6309163/c8sc01043a.pdf},
}

@article{LiNitzan2022,
   author = "Li, Tao E. and Cui, Bingyu and Subotnik, Joseph E. and Nitzan, Abraham",
   title = "Molecular Polaritonics: Chemical Dynamics Under Strong Light–Matter Coupling", 
   journal= "Annual Review of Physical Chemistry",
   year = "2022",
   volume = "73",
   number = "Volume 73, 2022",
   pages = "43-71",
   doi = "https://doi.org/10.1146/annurev-physchem-090519-042621",
   url = "https://www.annualreviews.org/content/journals/10.1146/annurev-physchem-090519-042621",
   publisher = "Annual Reviews",
   issn = "1545-1593",
   type = "Journal Article",
}

@article{Pandya2021,
author = {Pandya, Raj and Ashoka, Arjun and Georgiou, Kyriacos and Sung, Jooyoung and Jayaprakash, Rahul and Renken, Scott and Gai, Lizhi and Shen, Zhen and Rao, Akshay and Musser, Andrew J.},
title = {Tuning the Coherent Propagation of Organic Exciton-Polaritons through Dark State Delocalization},
journal = {Advanced Science},
volume = {9},
number = {18},
pages = {2105569},
doi = {https://doi.org/10.1002/advs.202105569},
url = {https://advanced.onlinelibrary.wiley.com/doi/abs/10.1002/advs.202105569},
eprint = {https://advanced.onlinelibrary.wiley.com/doi/pdf/10.1002/advs.202105569},
year = {2022}
}

@article{Weisbuch1992,
  title = {Observation of the coupled exciton-photon mode splitting in a semiconductor quantum microcavity},
  author = {Weisbuch, C. and Nishioka, M. and Ishikawa, A. and Arakawa, Y.},
  journal = {Phys. Rev. Lett.},
  volume = {69},
  issue = {23},
  pages = {3314--3317},
  numpages = {0},
  year = {1992},
  month = {Dec},
  publisher = {American Physical Society},
  doi = {10.1103/PhysRevLett.69.3314},
  url = {https://link.aps.org/doi/10.1103/PhysRevLett.69.3314}
}

@article{Hutchison2012,
author = {Hutchison, James A. and Schwartz, Tal and Genet, Cyriaque and Devaux, Eloïse and Ebbesen, Thomas W.},
title = {Modifying Chemical Landscapes by Coupling to Vacuum Fields},
journal = {Angewandte Chemie International Edition},
volume = {51},
number = {7},
pages = {1592-1596},
doi = {https://doi.org/10.1002/anie.201107033},
url = {https://onlinelibrary.wiley.com/doi/abs/10.1002/anie.201107033},
eprint = {https://onlinelibrary.wiley.com/doi/pdf/10.1002/anie.201107033},
year = {2012}
}

@article{Orgiu2015,
  author  = {Orgiu, E. and George, J. and Hutchison, J. A. and Devaux, E. and Dayen, J. F. and Doudin, B. and Stellacci, F. and Genet, C. and Schachenmayer, J. and Genes, C. and Pupillo, G. and Samor{\`i}, P. and Ebbesen, T. W.},
  title   = {Conductivity in Organic Semiconductors Hybridized with the Vacuum Field},
  journal = {Nature Materials},
  volume  = {14},
  pages   = {1123--1129},
  year    = {2015},
  doi     = {https://doi.org/10.1038/nmat4392},
}

@article{Thomas2016,
author = {Thomas, Anoop and George, Jino and Shalabney, Atef and Dryzhakov, Marian and Varma, Sreejith J. and Moran, Joseph and Chervy, Thibault and Zhong, Xiaolan and Devaux, Eloïse and Genet, Cyriaque and Hutchison, James A. and Ebbesen, Thomas W.},
title = {Ground-State Chemical Reactivity under Vibrational Coupling to the Vacuum Electromagnetic Field},
journal = {Angewandte Chemie International Edition},
volume = {55},
number = {38},
pages = {11462-11466},
doi = {https://doi.org/10.1002/anie.201605504},
url = {https://onlinelibrary.wiley.com/doi/abs/10.1002/anie.201605504},
eprint = {https://onlinelibrary.wiley.com/doi/pdf/10.1002/anie.201605504},
year = {2016}
}

@article{Feist2018,
author = {Feist, Johannes and Galego, Javier and Garcia-Vidal, Francisco J.},
title = {Polaritonic Chemistry with Organic Molecules},
journal = {ACS Photonics},
volume = {5},
number = {1},
pages = {205-216},
year = {2018},
doi = {10.1021/acsphotonics.7b00680},

URL = { 
    
        https://doi.org/10.1021/acsphotonics.7b00680
    
    

},
eprint = { 
    
        https://doi.org/10.1021/acsphotonics.7b00680
    
    

}

}

@article{GarciaVidal2021,
author = {Francisco J. Garcia-Vidal  and Cristiano Ciuti  and Thomas W. Ebbesen },
title = {Manipulating matter by strong coupling to vacuum fields},
journal = {Science},
volume = {373},
number = {6551},
pages = {eabd0336},
year = {2021},
doi = {10.1126/science.abd0336},
URL = {https://www.science.org/doi/abs/10.1126/science.abd0336},
eprint = {https://www.science.org/doi/pdf/10.1126/science.abd0336}
}

@article{Knapp1984,
title = {Lineshapes of molecular aggregates, exchange narrowing and intersite correlation},
journal = {Chemical Physics},
volume = {85},
number = {1},
pages = {73-82},
year = {1984},
issn = {0301-0104},
doi = {https://doi.org/10.1016/S0301-0104(84)85174-5},
url = {https://www.sciencedirect.com/science/article/pii/S0301010484851745},
author = {E.W. Knapp}
}

@article{Bassler1993,
author = {Bässler, H.},
title = {Charge Transport in Disordered Organic Photoconductors a Monte Carlo Simulation Study},
journal = {physica status solidi (b)},
volume = {175},
number = {1},
pages = {15-56},
doi = {https://doi.org/10.1002/pssb.2221750102},
url = {https://onlinelibrary.wiley.com/doi/abs/10.1002/pssb.2221750102},
eprint = {https://onlinelibrary.wiley.com/doi/pdf/10.1002/pssb.2221750102},
year = {1993}
}

@article{Spano2010,
author = {Spano, Frank C.},
title = {The Spectral Signatures of Frenkel Polarons in H- and J-Aggregates},
journal = {Accounts of Chemical Research},
volume = {43},
number = {3},
pages = {429-439},
year = {2010},
doi = {10.1021/ar900233v},
    note ={PMID: 20014774},

URL = { 
    
        https://doi.org/10.1021/ar900233v
    
    

},
eprint = { 
    
        https://doi.org/10.1021/ar900233v
    
    

}

}

@article{HestandSpano2018,
author = {Hestand, Nicholas
J. and Spano, Frank C.},
title = {Expanded Theory of H- and J-Molecular Aggregates: The Effects of Vibronic Coupling and Intermolecular Charge Transfer},
journal = {Chemical Reviews},
volume = {118},
number = {15},
pages = {7069-7163},
year = {2018},
doi = {10.1021/acs.chemrev.7b00581},
    note ={PMID: 29664617},

URL = { 
    
        https://doi.org/10.1021/acs.chemrev.7b00581
    
    

},
eprint = { 
    
        https://doi.org/10.1021/acs.chemrev.7b00581
    
    

}

}

@book{Mukamel1995,
  author    = {Mukamel, Shaul},
  title     = {Principles of Nonlinear Optical Spectroscopy},
  publisher = {Oxford University Press},
  address   = {New York},
  year      = {1995},
  isbn      = {9780195092783}
}

@book{MayKuhn2011,
  author    = {May, Volkhard and K{\"u}hn, Oliver},
  title     = {Charge and Energy Transfer Dynamics in Molecular Systems},
  edition   = {3},
  publisher = {Wiley-VCH},
  address   = {Weinheim},
  year      = {2011},
  doi       = {10.1002/9783527633791}
}

@article{Litinskaya2006,
  author  = {Litinskaya, M. and Reineker, P.},
  title   = {Inhomogeneous Broadening of Polaritons in Organic Microcavities},
  journal = {Physical Review B},
  volume  = {74},
  pages   = {165320},
  year    = {2006},
  doi     = {10.1103/PhysRevB.74.165320}
}

@article{Houdre1996,
  title = {Vacuum-field Rabi splitting in the presence of inhomogeneous broadening: Resolution of a homogeneous linewidth in an inhomogeneously broadened system},
  author = {Houdr\'e, R. and Stanley, R. P. and Ilegems, M.},
  journal = {Phys. Rev. A},
  volume = {53},
  issue = {4},
  pages = {2711--2715},
  numpages = {0},
  year = {1996},
  month = {Apr},
  publisher = {American Physical Society},
  doi = {10.1103/PhysRevA.53.2711},
  url = {https://link.aps.org/doi/10.1103/PhysRevA.53.2711}
}

@article{Lev2022,
author = {Cohn, Bar and Sufrin, Shmuel and Basu, Arghyadeep and Chuntonov, Lev},
title = {Vibrational Polaritons in Disordered Molecular Ensembles},
journal = {The Journal of Physical Chemistry Letters},
volume = {13},
number = {35},
pages = {8369-8375},
year = {2022},
doi = {10.1021/acs.jpclett.2c02341},
    note ={PMID: 36043884},

URL = { 
    
        https://doi.org/10.1021/acs.jpclett.2c02341
    
    

},
eprint = { 
    
        https://doi.org/10.1021/acs.jpclett.2c02341
    
    

}

}

@article{Qiang2026Arxiv,
    author = {Li, Tianchu and Venkatesh, Pranay and Shi, Qiang and Montoya-Castillo, Andrés},
    title = {For Molecular
Polaritons, Disorder and Phonon Time
Scales Control the Activation of Dark States in the Thermodynamic
Limit},
    journal = {The Journal of Physical Chemistry Letters},
    year = {2026},
    month = {09},
    issn = {1948-7185},
    doi = {10.1021/acs.jpclett.6c02068},
    url = {https://doi.org/10.1021/acs.jpclett.6c02068},
    eprint = {https://pubs.acs.org/jpclcd/article-pdf/doi/10.1021/acs.jpclett.6c02068/68664982/acs.jpclett.6c02068.pdf},
}

@article{Arkajit2023Rev,
author = {Mandal, Arkajit and Taylor, Michael A.D. and Weight, Braden M. and Koessler, Eric R. and Li, Xinyang and Huo, Pengfei},
title = {Theoretical Advances in Polariton Chemistry and Molecular Cavity Quantum Electrodynamics},
journal = {Chemical Reviews},
volume = {123},
number = {16},
pages = {9786-9879},
year = {2023},
doi = {10.1021/acs.chemrev.2c00855},
    note ={PMID: 37552606},

URL = { 
    
        https://doi.org/10.1021/acs.chemrev.2c00855
    
    

},
eprint = { 
    
        https://doi.org/10.1021/acs.chemrev.2c00855
    
    

}

}

@article{Xiang2019,
  author       = {Xiang, B. and Ribeiro, R. F. and Li, Y. and Dunkelberger, A. D. and Simpkins, B. B. and Yuen-Zhou, J. and Xiong, W.},
  title        = {Manipulating optical nonlinearities of molecular polaritons by delocalization},
  journal      = {Science Advances},
  year         = {2019},
  volume       = {5},
  number       = {9},
  pages        = {eaax5196},
  doi          = {10.1126/sciadv.aax5196},
  pmid         = {31799402},
  pmcid        = {PMC6868677},
  issn         = {2375-2548},
  publisher    = {American Association for the Advancement of Science (AAAS)}
}

@article{Russo2024,
author = {Russo, Mattia and Georgiou, Kyriacos and Genco, Armando and De Liberato, Simone and Cerullo, Giulio and Lidzey, David G. and Othonos, Andreas and Maiuri, Margherita and Virgili, Tersilla},
title = {Direct Evidence of Ultrafast Energy Delocalization Between Optically Hybridized J-Aggregates in a Strongly Coupled Microcavity},
journal = {Advanced Optical Materials},
volume = {12},
number = {25},
pages = {2400821},
doi = {https://doi.org/10.1002/adom.202400821},
url = {https://advanced.onlinelibrary.wiley.com/doi/abs/10.1002/adom.202400821},
eprint = {https://advanced.onlinelibrary.wiley.com/doi/pdf/10.1002/adom.202400821},
year = {2024}
}

@article{Juan2024,
  title = {Collective polaritonic effects on chemical dynamics suppressed by disorder},
  author = {P\'erez-S\'anchez, Juan B. and Mellini, Federico and Giebink, Noel C. and Yuen-Zhou, Joel},
  journal = {Phys. Rev. Res.},
  volume = {6},
  issue = {1},
  pages = {013222},
  numpages = {9},
  year = {2024},
  month = {Feb},
  publisher = {American Physical Society},
  doi = {10.1103/PhysRevResearch.6.013222},
  url = {https://link.aps.org/doi/10.1103/PhysRevResearch.6.013222}
}

@article{Joel2024NanoPhotonics,
author = {Schwennicke, Kai and Giebink, Noel C. and Yuen-Zhou, Joel},
title = {Extracting accurate light–matter couplings from disordered polaritons},
journal = {Nanophotonics},
volume = {13},
number = {14},
pages = {2469-2478},
doi = {https://doi.org/10.1515/nanoph-2024-0049},
url = {https://onlinelibrary.wiley.com/doi/abs/10.1515/nanoph-2024-0049},
eprint = {https://onlinelibrary.wiley.com/doi/pdf/10.1515/nanoph-2024-0049},
year = {2024}
}

@article{Musser2024Expt,
author = {George, Aleesha and Geraghty, Trevor and Kelsey, Zahra and Mukherjee, Soham and Davidova, Gloria and Kim, Woojae and Musser, Andrew J},
title = {Controlling the Manifold of Polariton States Through Molecular Disorder},
journal = {Advanced Optical Materials},
volume = {12},
number = {11},
pages = {2302387},
doi = {https://doi.org/10.1002/adom.202302387},
url = {https://advanced.onlinelibrary.wiley.com/doi/abs/10.1002/adom.202302387},
eprint = {https://advanced.onlinelibrary.wiley.com/doi/pdf/10.1002/adom.202302387},
year = {2024}
}

@article{Torma_2015,
doi = {10.1088/0034-4885/78/1/013901},
url = {https://doi.org/10.1088/0034-4885/78/1/013901},
year = {2015},
publisher = {IOP Publishing},
volume = {78},
number = {1},
pages = {012901},
author = {Törmä, P and Barnes, W L},
title = {Strong coupling between surface plasmon polaritons and emitters: a review},
journal = {Reports on Progress in Physics}
}

@article{Rider2022,
author = {Rider, Marie S. and Arul, Rakesh and Baumberg, Jeremy J. and Barnes, William L.},
title = {Theory of strong coupling between molecules and surface plasmons on a grating},
journal = {Nanophotonics},
volume = {11},
number = {16},
pages = {3695-3708},
doi = {https://doi.org/10.1515/nanoph-2022-0301},
url = {https://onlinelibrary.wiley.com/doi/abs/10.1515/nanoph-2022-0301},
eprint = {https://onlinelibrary.wiley.com/doi/pdf/10.1515/nanoph-2022-0301},
year = {2022}
}

@article{Daniel2021Chem,
title = {Understanding radiative transitions and relaxation pathways in plexcitons},
journal = {Chem},
volume = {7},
number = {4},
pages = {1092-1107},
year = {2021},
issn = {2451-9294},
doi = {https://doi.org/10.1016/j.chempr.2021.02.028},
url = {https://www.sciencedirect.com/science/article/pii/S2451929421001091},
author = {Daniel Finkelstein-Shapiro and Pierre-Adrien Mante and Sema Sarisozen and Lukas Wittenbecher and Iulia Minda and Sinan Balci and Tõnu Pullerits and Donatas Zigmantas}
}

@article{Cho2005JPCB,
author = {Cho, Minhaeng and Vaswani, Harsha M. and Brixner, Tobias and Stenger, Jens and Fleming, Graham R.},
title = {Exciton Analysis in 2D Electronic Spectroscopy},
journal = {The Journal of Physical Chemistry B},
volume = {109},
number = {21},
pages = {10542-10556},
year = {2005},
doi = {10.1021/jp050788d},
    note ={PMID: 16852278},

URL = { 
    
        https://doi.org/10.1021/jp050788d
    
    

},
eprint = { 
    
        https://doi.org/10.1021/jp050788d
    
    

}

}

@article{Cho2008ChemRev,
author = {Cho, Minhaeng},
title = {Coherent Two-Dimensional Optical Spectroscopy},
journal = {Chemical Reviews},
volume = {108},
number = {4},
pages = {1331-1418},
year = {2008},
doi = {10.1021/cr078377b},
    note ={PMID: 18363410},

URL = { 
    
        https://doi.org/10.1021/cr078377b
    
    

},
eprint = { 
    
        https://doi.org/10.1021/cr078377b
    
    

}

}

@article{Zeb2022PRA,
  title = {Analytical solution of the disordered Tavis-Cummings model and its Fano resonances},
  author = {Zeb, M. Ahsan},
  journal = {Phys. Rev. A},
  volume = {106},
  issue = {6},
  pages = {063720},
  numpages = {12},
  year = {2022},
  month = {Dec},
  publisher = {American Physical Society},
  doi = {10.1103/PhysRevA.106.063720},
  url = {https://link.aps.org/doi/10.1103/PhysRevA.106.063720}
}

@article{Engelhardt2023PRL,
  title = {Polariton Localization and Dispersion Properties of Disordered Quantum Emitters in Multimode Microcavities},
  author = {Engelhardt, Georg and Cao, Jianshu},
  journal = {Phys. Rev. Lett.},
  volume = {130},
  issue = {21},
  pages = {213602},
  numpages = {7},
  year = {2023},
  month = {May},
  publisher = {American Physical Society},
  doi = {10.1103/PhysRevLett.130.213602},
  url = {https://link.aps.org/doi/10.1103/PhysRevLett.130.213602}
}

@article{Aroeira2025Arxiv,
    author = {Aroeira, Gustavo J. R. and Ribeiro, Raphael F.},
    title = {Static disorder-induced renormalization of polariton group velocity},
    journal = {The Journal of Chemical Physics},
    volume = {163},
    number = {12},
    pages = {124120},
    year = {2025},
    month = {09},
    issn = {0021-9606},
    doi = {10.1063/5.0288551},
    url = {https://doi.org/10.1063/5.0288551},
    eprint = {https://pubs.aip.org/aip/jcp/article-pdf/doi/10.1063/5.0288551/20713363/124120_1_5.0288551.pdf},
}

@article{Wu2022JCP,
    author = {Wu, Weijun and Sifain, Andrew E. and Delpo, Courtney A. and Scholes, Gregory D.},
    title = {Polariton enhanced free charge carrier generation in donor–acceptor cavity systems by a second-hybridization mechanism},
    journal = {The Journal of Chemical Physics},
    volume = {157},
    number = {16},
    pages = {161102},
    year = {2022},
    month = {10},
    issn = {0021-9606},
    doi = {10.1063/5.0122497},
    url = {https://doi.org/10.1063/5.0122497},
    eprint = {https://pubs.aip.org/aip/jcp/article-pdf/doi/10.1063/5.0122497/16552152/161102_1_online.pdf},
}

@misc{Wu2024Arxiv,
      title={Disorder enhanced exciton transport and quantum information spreading with the assistance of cavity QED}, 
      author={Weijun Wu and Ava N. Hejazi and Gregory D. Scholes},
      journal= {arXiv preprint arXiv:2409.06900},
      year={2024},
      archivePrefix={arXiv},
      primaryClass={quant-ph},
      url={https://arxiv.org/abs/2409.06900},
      eprint={2409.06900},
}

@article{Catuto2025JCP,
    author = {Catuto, Robert F. and Chen, Hsing-Ta},
    title = {Interplay between static and dynamic disorder: Contrasting effects on dark state population inside a cavity},
    journal = {The Journal of Chemical Physics},
    volume = {162},
    number = {20},
    pages = {204102},
    year = {2025},
    month = {05},
    issn = {0021-9606},
    doi = {10.1063/5.0270892},
    url = {https://doi.org/10.1063/5.0270892},
    eprint = {https://pubs.aip.org/aip/jcp/article-pdf/doi/10.1063/5.0270892/20528921/204102_1_5.0270892.pdf},
}

@article{Li2025JPCL,
    author = {Li, Wei-Kuo and Chen, Hsing-Ta},
    title = {Disorder-Induced
Spectral Splitting versus Rabi Splitting
under Strong Light-Matter Coupling},
    journal = {The Journal of Physical Chemistry Letters},
    volume = {16},
    number = {26},
    pages = {6728-6733},
    year = {2025},
    month = {06},
    issn = {1948-7185},
    doi = {10.1021/acs.jpclett.5c01302},
    url = {https://doi.org/10.1021/acs.jpclett.5c01302},
    eprint = {https://pubs.acs.org/jpclcd/article-pdf/16/26/6728/42064289/jz5c01302.pdf},
}

@article{Sidler2021JPCL,
    author = {Sidler, Dominik and Sch\"afer, Christian and Ruggenthaler, Michael and Rubio, Angel},
    title = {Polaritonic Chemistry: Collective Strong Coupling
Implies Strong Local Modification of Chemical Properties},
    journal = {The Journal of Physical Chemistry Letters},
    volume = {12},
    number = {1},
    pages = {508-516},
    year = {2021},
    month = {01},
    issn = {1948-7185},
    doi = {10.1021/acs.jpclett.0c03436},
    url = {https://doi.org/10.1021/acs.jpclett.0c03436},
    eprint = {https://pubs.acs.org/jpclcd/article-pdf/12/1/508/67602630/acs.jpclett.0c03436.pdf},
}

@article{Sidler2024JPCL,
    author = {Sidler, Dominik and Schnappinger, Thomas and Obzhirov, Anatoly and Ruggenthaler, Michael and Kowalewski, Markus and Rubio, Angel},
    title = {Unraveling a
Cavity-Induced Molecular Polarization
Mechanism from Collective Vibrational Strong Coupling},
    journal = {The Journal of Physical Chemistry Letters},
    volume = {15},
    number = {19},
    pages = {5208-5214},
    year = {2024},
    month = {05},
    issn = {1948-7185},
    doi = {10.1021/acs.jpclett.4c00913},
    url = {https://doi.org/10.1021/acs.jpclett.4c00913},
    eprint = {https://pubs.acs.org/jpclcd/article-pdf/15/19/5208/3374526/jz4c00913.pdf},
}

@article{Horak2025PRR,
  title = {Analytic model reveals local molecular polarizability changes induced by collective strong coupling in optical cavities},
  author = {Horak, Jacob and Sidler, Dominik and Schnappinger, Thomas and Huang, Wei-Ming and Ruggenthaler, Michael and Rubio, Angel},
  journal = {Phys. Rev. Res.},
  volume = {7},
  issue = {1},
  pages = {013242},
  numpages = {15},
  year = {2025},
  month = {Mar},
  publisher = {American Physical Society},
  doi = {10.1103/PhysRevResearch.7.013242},
  url = {https://link.aps.org/doi/10.1103/PhysRevResearch.7.013242}
}

@article{Castagnola2025PRX,
  title = {Realistic Ab Initio Predictions of Excimer Behavior under Collective Light-Matter Strong Coupling},
  author = {Castagnola, Matteo and Lexander, Marcus T. and Koch, Henrik},
  journal = {Phys. Rev. X},
  volume = {15},
  issue = {2},
  pages = {021040},
  numpages = {17},
  year = {2025},
  month = {May},
  publisher = {American Physical Society},
  doi = {10.1103/PhysRevX.15.021040},
  url = {https://link.aps.org/doi/10.1103/PhysRevX.15.021040}
}

@article{Dutta2024ThermalDisorder,
  author  = {Dutta, Arpan and Tiainen, Ville and Sokolovskii, Ilia and Duarte, Lu{\'\i}s and Marke{\v{s}}evi{\'c}, Nemanja and Morozov, Dmitry and Qureshi, Hassan A. and Pikker, Siim and Groenhof, Gerrit and Toppari, J. Jussi},
  title   = {Thermal disorder prevents the suppression of ultra-fast photochemistry in the strong light-matter coupling regime},
  journal = {Nature Communications},
  volume  = {15},
  pages   = {6600},
  year    = {2024},
  doi     = {10.1038/s41467-024-50532-5},
  url     = {https://doi.org/10.1038/s41467-024-50532-5}
}

@article{KochRubioJCP2021AbInitioStudy,
    author = {Haugland, Tor S. and Schäfer, Christian and Ronca, Enrico and Rubio, Angel and Koch, Henrik},
    title = {Intermolecular interactions in optical cavities: An ab initio QED study},
    journal = {The Journal of Chemical Physics},
    volume = {154},
    number = {9},
    pages = {094113},
    year = {2021},
    month = {03},
    issn = {0021-9606},
    doi = {10.1063/5.0039256},
    url = {https://doi.org/10.1063/5.0039256},
    eprint = {https://pubs.aip.org/aip/jcp/article-pdf/doi/10.1063/5.0039256/20638670/094113_1_5.0039256.pdf},
}

@article{Biswas2025ESC,
  author  = {Biswas, Subha and Mondal, Mainak and Chandrasekharan, Gokul and Mony, Kavya S. and Singh, Akshay and Thomas, Anoop},
  title   = {Electronic strong coupling modifies the ground-state intermolecular interactions in self-assembled chlorin molecules},
  journal = {Nature Communications},
  volume  = {16},
  pages   = {5115},
  year    = {2025},
  doi     = {10.1038/s41467-025-60025-8},
  url     = {https://doi.org/10.1038/s41467-025-60025-8}
}

@article{Ebbesen2024VSCinNMR,
  author  = {Patrahau, B. and Piejko, M. and Mayer, R. J. and Antheaume, C. and Sangchai, T. and Ragazzon, G. and Jayachandran, A. and Devaux, E. and Genet, C. and Moran, J. and Ebbesen, T. W.},
  title   = {Direct Observation of Polaritonic Chemistry by Nuclear Magnetic Resonance Spectroscopy},
  journal = {Angewandte Chemie International Edition},
  volume  = {63},
  number  = {23},
  pages   = {e202401368},
  year    = {2024},
  doi     = {10.1002/anie.202401368},
  url     = {https://doi.org/10.1002/anie.202401368}
}

\end{document}